\pdfoutput=1
\documentclass[11pt]{article}
\usepackage[svgnames,dvipsnames]{xcolor}
\usepackage[height=8.85in,width=6.45in]{geometry}
\usepackage{here, amsmath, amsthm, latexsym, amssymb, bm, ascmac, mathtools, multicol, subfig, url, overpic, braket, tikz, cite, xcolor}
\numberwithin{equation}{section}
\theoremstyle{plain}

\newtheorem*{thm*}{Theorem}

\newtheorem*{claim*}{Claim}
\usepackage[title,titletoc]{appendix}
\usepackage[pdfencoding=auto,pagebackref]{hyperref}
\definecolor{Myblue}{rgb}{0,0,0.9}
\hypersetup{
    linktocpage=true,
    colorlinks=true,
    citecolor=Myblue,
    linkcolor=Myblue,
    urlcolor=Myblue,
}

\begin{document}
\title{\bf Gauging on the lattice and gapped/gapless topological phases}
\date{\today}
\author{Takamasa Ando \\ 
\textit{Center for Gravitational Physics and Quantum Information,}\\
\textit{Yukawa Institute for Theoretical Physics, Kyoto University, 
Kyoto 606-8502, Japan}}
\maketitle
\begin{abstract}  
    In this work, we explore topological phases of matter obtained by \textit{effectively} gauging or fermionizing a system, where the Gauss law constraint is only enforced energetically. In contrast to conventional gauging or fermionization, the symmetry that is effectively gauged at low energies still generates a global symmetry that acts on the whole Hilbert space faithfully. This symmetry turns out to protect a nontrivial topological phase together with other symmetries, or it can carry a nontrivial emergent 't Hooft anomaly. We provide a precise formula for the topological response action involving these symmetries in a general setup, as well as a formula for 't Hooft anomalies. As an application, we apply the general treatment of the procedure to gapless systems and find various new gapless SPT phases, such as the one carrying the Gu-Wen fermionic anomalies at low energy.
\end{abstract}

\setcounter{tocdepth}{2}
\tableofcontents

\section{Introduction and summary}\label{sec_intro}
Consider a bosonic theory \(\mathsf{D}\) in \(d\)-spacetime dimensions with a global zero-form \(\Gamma\) symmetry.\footnote{In this paper, we only consider invertible symmetries, i.e., each element of the symmetry category has an inverse element.} Here, \(\Gamma\) is a finite group, which fits into the following central extension.
\begin{equation}\label{extension_intro}
    1 \rightarrow A \rightarrow \Gamma \rightarrow G \rightarrow 1.
\end{equation}
Here, \(A\) must be abelian, and an element of the second group cohomology group \([e]\in H^2(G,A)\) specifies an isomorphic class of this central extension. In this paper, we discuss the gauging procedure on the lattice. Specifically, we start from the local lattice Hamiltonian of the form
\begin{equation}
    H_{\mathsf{D}}=\sum_{j}h_{j},
\end{equation}
where \(h_{j}\) is an interaction term localized at site \(j\) and has \(A\) symmetry generated by on-site operators. Note that such on-site symmetries are always non-anomalous and we can gauge such non-anomalous symmetries.
To implement the gauging, we introduce dual fields (gauge fields) on links so that the global \(A\) symmetry acts trivially on the gauged theory. We assume that dual fields are minimally coupled. Denoting the gauged interaction term by \(h_{j}^g\), we have the gauged Hamiltonian of the form  
\begin{equation}
    H_{\mathsf{D}/A}=\sum_{j}h_{j}^g.
\end{equation}
In addition to this modification, we need to impose a non-local constraint on the system, called the Gauss law constraint. The constraint is defined so that any state of the system must be invariant under the actions of the Gauss law operators \(\{G_{j}\}_{j}\). After gauging, the original \(A\) symmetry is no longer a global symmetry, but a local symmetry. The global symmetry of the gauged system is a mixture of \((d-2)\)-form \(\hat{A}=\operatorname{Hom}(A,U(1))\cong A\) symmetry and zero-form ``\(G\) symmetry''.\footnote{The full structure of global symmetries of the gauged system depends on the extension class and anomalies of the original system. Whereas the \(G\) symmetry survives under gauging when the original system is non-anomalous, the surviving \(G\) defects in an anomalous system do not necessarily form a \(G\) group structure; see \cite{Tachikawa:2017gyf,Kaidi:2021xfk}.}

For s system with a \(\mathbb{Z}_2\) \((d-2)\)-form symmetry with \(Sq^2\) anomaly, we can also define fermionization, which maps a bosonic system to a fermionic system. We review the fermionization scheme in Sec.~\ref{sec_fermionization}. After fermionization, we obtain the fermionic Hamiltonian of the form
\begin{equation}\label{fermi_ham_intro}
    H_{\mathcal{F}(\mathsf{D})}=\sum_{j}h_{j}^f
\end{equation}
with the Gauss law constraint. The fermionized model \eqref{fermi_ham_intro} has the fermion parity symmetry \(\mathbb{Z}_2^F\). In this paper, we mainly consider fermionization in \((1+1)\) dimensions, where the situations fit into \eqref{extension_intro}.
\bigskip

In the field theory formalism, the partition functions of the gauged and fermionized theories with strict Gauss law constraints are given by 
\begin{gather}
    Z_{\mathsf{D}/A}[\hat{A},G;e]=\#\sum_{a}Z_{\mathsf{D}}[G,a;e]\,e^{2\pi i\int (a,\hat{A})},\\
    Z_{\mathsf{F}}[\rho,G;e]=\#\sum_{a}Z_{\mathsf{D}}[G,a;e]\,z(\rho,a;b),\label{usual_fer_qft}
\end{gather}
where \([e]\in H^2(G,A)\) specifies the extension class. \(\hat{A}\) is a background gauge field for the dual global symmetry and \(G\) is a background gauge field for \(G\) symmetry. \((a,\hat{A})\) denotes the pairing of the dynamical gauge field \(a\) and \(\hat{A}\). \(\rho\) specifies the spin structure of the spacetime manifold, i.e.~\(\delta\rho=w_{2}\), where \(w_{2}\) is the second Stiefel-Whitney class of the spacetime manifold. \(z(\rho,a;b)\) is some fermionic generalization of discrete torsion, and we detail the definition in Sec.~\ref{sec_fermionQFT}. The prefactor \(\#\) depends on the conventions and is determined up to an arbitrary power of Euler counterterms. For \(d=1+1\) and \([e]=0\), \(z(\rho,a;b)\) is explicitly given as
\begin{equation}\label{arf_intro}
    z(\rho,a;b=0)=(-1)^{\operatorname{Arf}(a+\rho)+\operatorname{Arf}(\rho)}.
\end{equation}
Note that the factors \(e^{2\pi i\int (a,\hat{A})}\) and \(z(\rho,a;b)\) give anomalies in general, and have phase ambiguities under the gauge transformation of background gauge fields \cite{Tachikawa:2017gyf}.
\bigskip

Instead of imposing strict Gauss law constraints, in this paper we explore the following Hamiltonian, where the Gauss law is only enforced energetically:
\begin{equation}\label{gaugeHam}
    H_{\mathsf{D}/A}^K=\sum_{j}h_j^g-K\sum_{j}(G_{j}+G_{j}^\dagger),
\end{equation}
where \(K\) is a sufficiently large positive constant. Since the Gauss law operators \(\{G_{j}\}_{j}\) commute with \(h_{j}^g\) and they commute with each other, the ground states of this Hamiltonian satisfy the Gauss law constraint. In particular, we recover the Gauss law in the limit of \(K\) to infinity. Similarly, for fermionization we consider 
\begin{equation}\label{gaugeHamF}
    H_{\mathcal{F}(\mathsf{D})}^K=\sum_{j}h_j^f-K\sum_{j}(G_{j}^f+(G_{j}^{f})^\dagger),
\end{equation}
where \(G_{j}^f\) is the Gauss law operator for fermionization. In this paper we will refer to the procedure to obtain the Hamiltonian \eqref{gaugeHam}/\eqref{gaugeHamF} as \emph{effective gauging/fermionization}. Enforcing the Gauss law only energetically was previously introduced in the literature e.g.~\cite{Borla:2020avq,Verresen:2022mcr}. In this paper, we extend this idea and discuss effective gauging/fermionization in general setups. 
Note that symmetry operators for the original global symmetry \(\Gamma\) act faithfully on the Hilbert space. Namely, global symmetries of \eqref{gaugeHam}, \eqref{gaugeHamF} are \(\hat{A}\times \Gamma\), \(\mathbb{Z}_2^F\times \Gamma\), respectively. In the field theory formalism, the partition functions of \eqref{gaugeHam} and \eqref{gaugeHamF} should contain the background gauge field for the \(A\) symmetry. Therefore, we need to specify the precise topological response actions for all \(\hat{A}\), \(G\), and \(A\), and that is the main objective of this paper. 
\bigskip

In this setup, the main claim of the paper is as follows.
\begin{claim*}
    Hamiltonian defined by \eqref{gaugeHam} belongs to a topological phase with symmetry \(\Gamma\times \hat{A}\), whose partition function is
    \begin{equation}\label{eff_response}
        Z_{\mathsf{D}/A}[\hat{A},G,A;e]=Z_{\mathsf{D}/A}[\hat{A},G;e]\,e^{-2\pi i\int(A,\hat{A})}=\#\sum_{a}Z_{\mathsf{D}}[G,a]\,e^{2\pi i\int (a-A,\hat{A})}.
    \end{equation}
    For fermionization, the Hamiltonian \eqref{gaugeHamF} is described by 
    \begin{equation}
        Z_{\mathsf{F}}[\rho,G,A;e]=\#\sum_{a}Z_{\mathsf{D}}[G,a;e]\,z(\rho,a+A;b).
    \end{equation}
\end{claim*}
We discuss these response actions in terms of both the lattice model analysis and the field theory formalism.

\subsection*{Application to gapped and gapless topological phases}
For unique gapped systems, we have a class of topological phases, called symmetry protected topological (SPT) phases \cite{Pollmann:2009mhk,Pollmann:2009ryx,Chen:2010zpc,Schuch:2010,Chen:2011pg,Levin:2012yb}. Topological response actions of SPTs are given by some well-defined \(U(1)\)-valued functions of background gauge fields for global symmetries. Suppose that the extension class is trivial and the system is non-anomalous. Then the additional phase \(e^{-2\pi i\int(A,\hat{A})}\) in \eqref{eff_response} is a partition function of an SPT phase. Therefore, we can construct various lattice Hamiltonians stacked by such SPT phases. In general, this phase factor carries 't Hooft anomalies, which cancel the ones from \(e^{2\pi i\int (a,\hat{A})}\).

An intriguing application of effective gauging and fermionization is to construct lattice models for gapless topological phases, which have been the subject of recent intensive studies \cite{Scaffidi:2017ppg,Verresen:2019igf,Thorngren:2020wet,Borla:2020avq,Ma:2021dfx,Yu:2021rng,Li:2022jbf,Wen:2022tkg,Hidaka:2022bbz,Li:2023knf,Wen:2023otf,Huang:2023pyk,Yu:2024toh}. Non-anomalous gapless theories with emergent anomalies are called intrinsically gapless SPTs (igSPTs). In our construction \eqref{gaugeHam} and \eqref{gaugeHamF}, igSPT models are obtained from gapless Hamiltonians and non-anomalous global symmetries, which are nontrivially extended by \(A\). Since the choice of input non-anomalous gapless models is arbitrary, we can systematically construct igSPT models. On the other hand, not all anomalies are obtained by gauging some non-anomalous theories. Nevertheless, for wide classes of anomalies we can construct intrinsically gapless SPT models with such emergent anomalies by combining effective gauging and condensations. We will use this technique to construct an igSPT model with Gu-Wen fermionic anomalies, which are realized as boundary theories of Gu-Wen fermionic SPT phases \cite{Gu:2012ib}.

\paragraph{Organization of the paper:}
The rest of the paper is organized as follows. In Sec.~\ref{sec_gauge_anomaly}, we review \(\mathbb{Z}_2\) gauging and fermionization in \((1+1)\) dimensions on the lattice. In Sec.~\ref{sec_emergent_dual}, we introduce lattice models in which we enforce the Gauss law only energetically. We claim that the original global symmetries are also global symmetries in the models and provide additional topological response action with these symmetries. As an application, we give some lattice models for both gapped and gapless topological phases. In Sec.~\ref{sec_fermion}, we continue the discussions on fermionic theories. In Sec.~\ref{sec_highdim}, we briefly provide comments on generalizations of constructions of gapped and gapless models to higher dimensions. In Sec.~\ref{sec_QFT}, we discuss the topics based on the quantum field theory (QFT) formalism. Readers may read Sec.~\ref{sec_QFT} and other sections separately.

\paragraph{Notations:}
We always take the periodic boundary condition for \((1+1)\)-dimensional lattice models. This assumption is not essential and we can consider appropriate boundary conditions for given global symmetries.

\section{Gauging and anomaly on the lattice}\label{sec_gauge_anomaly}
In this section, we review the gauging and fermionization procedure on \((1+1)\)-dimensional lattice. Though we use some concrete lattice models for the explanation, gauging and fermionization can be applied to generic lattice models with on-site symmetries. We also review the way to detect anomalies on the lattice. 
\subsection{Gauging \texorpdfstring{\(\mathbb{Z}_2\)}{Z2} symmetry}
Consider a one-dimensional quantum spin chain of size \(L\) with a \(\mathbb{Z}_2\) global symmetry generated by
\begin{equation}
    U_{\mathbb{Z}_2}=\prod_{j=1}^{L}\sigma_{j}^x,
\end{equation}
where \(\sigma_{j}^\alpha\,(\alpha=x,y,z)\) are the Pauli matrices on the site \(j\). Each local Hilbert space \(\mathcal{H}_{j}\) is isomorphic to \(\mathbb{C}^2\) and the total Hilbert space is \(\mathcal{H}=\bigotimes_{j}\mathcal{H}_{j}\). An example of a lattice model for this \(\mathbb{Z}_2\) symmetry is the transverse-field Ising (TFI) model defined by
\begin{equation}\label{TFI_ham}
    H_{\text{TFI}}=-\sum_{j=1}^L \left(\sigma_{j}^x+J\sigma_j^z\sigma_{j+1}^z\right),
\end{equation}
where \(J\) is a positive constant. 

Since the global symmetry is on-site, we can gauge the \(\mathbb{Z}_2\) global symmetry.\footnote{Note that on-site symmetries (also without any projective representation) admit unique gapped ground states. At the level of field theories, symmetries that admit unique gapped ground states are gaugeable.} To implement gauging, we introduce the dual fields \(\{\tau_{j+1/2}^\alpha\}_{j}\) on the link and couple them to the Hamiltonian as follows.
\begin{equation}
    H_{\text{TFI}}^g=-\sum_{j=1}^{L}\left(\sigma_{j}^x+J\sigma_j^z\tau_{j+1/2}^x\sigma_{j+1}^z\right).
\end{equation}
The local interactions in \(H_{\text{TFI}}^g\) must commute with the Gauss law operators \(\{G_{j}\}_{j}\). Here, each \(G_{j}\) generates the local gauge transformations: \(\sigma_{j}^z\mapsto -\sigma_{j}^z,\tau_{j\pm 1/2}^x\mapsto -\tau_{j\pm 1/2}^x\). Such operators are given by \(G_{j}\coloneqq \tau_{j-1/2}^z\sigma_{j}^x\tau_{j+1/2}^z\). Moreover, the physical state \(\ket{\psi}\) should be invariant under the action of \(G_{j}\) because of the gauge principle. Namely, the physical Hilbert space \(\mathcal{H}^{\text{phys}}\) of the gauged theory is 
\begin{equation}
    \mathcal{H}^{\text{phys}}=\left\{\ket{\psi}\in\mathcal{H}\otimes\mathcal{H}^g\mid G_j\ket{\psi}=\ket{\psi},\,\text{for arbitrary } j\right\},
\end{equation}
where \(\mathcal{H}^g\) is the Hilbert space of dual fields and isomorphic to \(\mathcal{H}\) on closed chains. We see that \(\operatorname{dim}_{\mathbb{C}}(\mathcal{H}^{\text{phys}})=2^L\). We can choose the basis so that only dual fields appear in the Hamiltonian. The unitary operator for mapping to such a basis is given by
\begin{equation}\label{CZ_def}
    V\coloneqq \prod_{j=1}^L e^{\frac{\pi i}{4}(1-\sigma_{j}^z)(1-\tau_{j-1/2}^z)}e^{\frac{\pi i}{4}(1-\sigma_{j}^z)(1-\tau_{j+1/2}^z)}.
\end{equation}
Then we obtain 
\begin{equation}
    \widetilde{H}_{\text{TFI}}^g=VH_{\text{TFI}}^gV^{-1}=-\sum_{j=1}^L\left(\tau_{j-1/2}^z\sigma_{j}^x\tau_{j+1/2}^z+J\tau_{j+1/2}^x\right)
\end{equation}
Under this unitary transformation, the Gauss law operators are also changed as \(G_{j}\mapsto VG_{j}V^\dagger=\sigma_{j}^x\). Therefore, in this basis, the gauged Hamiltonian is equivalent to 
\begin{equation}
    \widetilde{H}_{\text{TFI}}^g\simeq{\widetilde{H}_{\text{TFI}}^{g^\prime}}=-\sum_{j=1}^L\left(\tau_{j-1/2}^z\tau_{j+1/2}^z+J\tau_{j+1/2}^x\right).
\end{equation}
We no longer have any non-local constraints in this expression. The transmutation defined by \(\sigma_{j}^z\mapsto\tau_{j-1/2}^z\tau_{j+1/2}^z,\sigma_{j}^z\sigma_{j+1}^z\mapsto\tau_{j+1/2}^x\) is the Kramers-Wannier transformation.

\subsection{Fermionization}\label{sec_fermionization}
\subsubsection{Jordan-Wigner transformation}
Consider a \((1+1)\)-dimensional fermionic system \(\mathsf{F}\). On each site \(j\), we have two Majorana operators \(\gamma_j,\gamma_j^\prime\).\footnote{Majorana operators satisfy the anti-commutation relation \(\{\gamma_i,\gamma_j\}=\{\gamma_i^\prime,\gamma_j^\prime\}=2\delta_{i,j},\{\gamma_i,\gamma_j^\prime\}=0\).}
For fermionic systems, We always have a \(\mathbb{Z}_2\) global symmetry generated by the fermion parity operator
\begin{equation}
    (-1)^F\coloneqq \prod_{j=1}^L\left(-i\gamma_j\gamma_j^\prime\right).
\end{equation}
The boundary condition for Majorana operators is \(\gamma_{j+L}=-(-1)^{t_f}\gamma_j,\gamma_{j+L}^\prime=-(-1)^{t_f}\gamma_j^\prime\), where \(t_f=0,1\) correspond to the anti-periodic and the periodic boundary condition, respectively.
    
Jordan-Wigner (JW) transformation is a duality transformation that exchanges the two systems \(\mathsf{D}\) (\(\mathbb{Z}_2\) symmetric spin system) and \(\mathsf{F}\). The duality map is defined as
\begin{equation}
     \sigma^x_j=-i\gamma_j\gamma_j^\prime,\quad
    \sigma^z_j=\prod_{l<j}\left(-i\gamma_l\gamma_l^\prime\right)\gamma_j,\quad 
    \sigma^y_j=\prod_{l<j}\left(-i\gamma_l\gamma_l^\prime\right)\gamma_j^\prime,
\end{equation}
\begin{equation}
    \gamma_j=\left(\prod_{l<j}\sigma^x_l\right)\sigma^z_j,\quad 
    \gamma_j^\prime=-\left(\prod_{l<j}\sigma^x_l\right)\sigma^y_j.    
\end{equation}
This JW transformation can be defined both for closed chains and open chains \cite{Hsieh:2020uwb}. After implementing the JW transformation to the TFI Hamiltonian \eqref{TFI_ham}, we obtain 
\begin{equation}\label{JW_TFI}
    H_{\mathcal{JW}(\text{TFI})}=\sum_{j=1}^{L}\left(i\gamma_{j}\gamma_{j}^\prime+Ji\gamma_{j}^\prime\gamma_{j+1}\right).
\end{equation}
This model is known as the Kitaev chain \cite{Kitaev:2000nmw}. 

\subsubsection{Fermionization}
We also have another duality transformation to have fermionic systems from a non-anomalous \(\mathbb{Z}_2\) symmetric system. It is known as fermionization \cite{Karch:2019lnn,Ji:2019ugf}. The JW transformation and fermionization are related to each other through the following diagram.\footnote{The definition of fermionization differs across literature. Some people define fermionization to be Jordan-Wigner transformation or the corresponding manipulation in the field theory formalism.}
\begin{equation}\label{fer_com_dig}
    \begin{tikzpicture}
        \centering
        \pgfmathsetmacro{\sepx}{3}
        \pgfmathsetmacro{\sepy}{2}
        \node[] (D) at (0,0) {\(\mathsf{D}\)};
        \node[] (Fp) at (\sepx,0) {\(\mathsf{F}^\prime\)};
        \node[] (Dp) at (0,-\sepy) {\(\mathsf{D}^\prime\)};
        \node[] (F) at (\sepx,-\sepy) {\(\mathsf{F}\)};

        \draw[<->] (D) -- node[scale=.8,anchor=south] {Jordan-Wigner} (Fp);
        \draw[<->] (D) -- node[scale=.8,anchor=east] {\(\mathbb{Z}_2\) gauging} (Dp);
        \draw[<->] (Dp) -- node[scale=.8,anchor=north] {Jordan-Wigner} (F);
        \draw[<->] (Fp) -- node[scale=.8,anchor=west] {\(\times~\text{Kitaev}\)} (F);
        \draw[->,transform canvas={yshift=2pt}] (D) -- node[scale=.8,anchor=south,sloped] {Fermionization} (F);
        \draw[->,transform canvas={yshift=-2pt}] (F) -- node[scale=.8,anchor=north,sloped] {Bosonization} (D);
    \end{tikzpicture},
\end{equation}
where \(\times~\text{Kitaev}\) denotes ``stacking the Kitaev phase''. The Kitaev phase is the nontrivial invertible phase with the fermion parity symmetry. See e.g.~\cite{Hsieh:2020uwb} for the details of the diagram. From the definition, we immediately implement fermionization for the TFI model as follows:
\begin{equation}
    H_{\mathcal{F}(\text{TFI})}=-\sum_{j=1}^L \left(\sigma_{j}^x+J\sigma_j^z(-i\gamma_{j+1/2}\gamma_{j+1/2}^\prime)\sigma_{j+1}^z\right),
\end{equation}
and we also need to impose the Gauss law constraint on the physical state. The Gauss law operator \(G_{j}^f\) for fermionization is given by
\begin{equation}
    G_j^f=(-i\gamma_{j-1/2}^\prime\gamma_{j+1/2})\sigma_j^x.
\end{equation}
As in \(\mathbb{Z}_2\) gauging, we can change the basis so that only the dual fields \(\gamma_{j+1/2},\gamma_{j+1/2}^\prime\) appear. The unitary conjugation operator is given by
\begin{equation}\label{Vf}
    V_{f}=\prod_{(j-1/2,j,j+1/2)}e^{\frac{\pi i}{4}(1-\sigma_{j}^z)(1+i\gamma^\prime_{j-1/2}\gamma_{j+1/2})}.
\end{equation}
Then we obtain
\begin{equation}\label{F_TFI}
    \widetilde{H}_{\mathcal{F}(\text{TFI})}=V_{f}H_{\mathcal{F}(\text{TFI})}V_{f}^\dagger=-\sum_{j=1}^L \left(\sigma_{j}^x(-i\gamma_{j-1/2}^\prime\gamma_{j+1/2})+J(-i\gamma_{j+1/2}\gamma_{j+1/2}^\prime)\right).
\end{equation}
In this basis, the Gauss law operator is \(V_{f}G_{j}^fV_{f}^\dagger=\sigma_{j}^x\). Since the Gauss law fixes so that \(\sigma_{j}^x=1\) for all \(j\), the spectrum of  is the same as 
\begin{equation}\label{fer_TFI_ham}
    \widetilde{H^\prime}_{\mathcal{F}(\text{TFI})}=\sum_{j=1}^{L}\left(i\gamma_{j-1/2}^\prime\gamma_{j+1/2}+Ji\gamma_{j+1/2}\gamma_{j+1/2}^\prime\right).
\end{equation}
Two Hamiltonians \(H_{\mathcal{JW}(\text{TFI})}\) and \(\widetilde{H^\prime}_{\mathcal{F}(\text{TFI})}\) are related via half translations of the Majorana: \(\gamma_{j-1/2}\mapsto \gamma_{j-1/2}^\prime,\gamma_{j-1/2}^\prime\mapsto\gamma_{j+1/2}\), and this operation corresponds to ``staking the Kitaev phase''. To understand the phrase, note that two Hamiltonians 
\begin{equation}
    H_{\text{trivial}}=\sum_{j}i\gamma_{j+1/2}\gamma_{j+1/2}^\prime,\quad H_{\text{Kitaev}}=\sum_{j}i\gamma_{j-1/2}^\prime\gamma_{j+1/2}
\end{equation}
are fixed models for the trivial phase and the nontrivial phase of \((1+1)\)-dimensional fermionic invertible phases, and the half translation of Majorana operators exchange these two Hamiltonians.

One can also see the relation of two theories \(\mathsf{F}\) and \(\mathsf{F}^\prime\) more explicitly. Specifically, let us start with the TFI Hamiltonian \(H_{\text{TFI}}\). After the Kramers-Wannier transformation, we obtain 
\begin{equation}
    H_{\mathcal{KW}(\text{TFI})}=-\sum_{j=1}^{L}\left(\sigma_{j-1}^z\sigma_{j}^z+J\sigma_{j}^x\right).
\end{equation} 
Then Jordan-Wigner transformation maps it to 
\begin{equation}\label{JWKW_TFI}
    H_{\mathcal{JW}(\mathcal{KW}(\text{TFI}))}=\sum_{j=1}^{L}\left(i\gamma_{j-1}^\prime\gamma_{j}+Ji\gamma_{j}\gamma_{j}^\prime\right).
\end{equation}
This is the same as the fermionized Hamiltonian \eqref{fer_TFI_ham} and we see \eqref{JW_TFI} and \eqref{JWKW_TFI} are related via the half Majorana translation.

\subsection{Else-Nayak procedure}
Once one obtains the expressions of symmetry operators in \((1+1)\)-dimensional systems, one can extract elements of \(H^3(G,U(1))\), which specify the anomaly of the system \cite{Else:2014vma,Kawagoe:2021gqi, Seifnashri:2023dpa}. In this paper, we use the procedure developed by Else and Nayak \cite{Else:2014vma}. We review this procedure in this subsection.

Let the system have a \(G\) global symmetry and \(U(g)\) be the unitary symmetry operators, which are not necessarily on-site operators. On the closed chain, \(U(g)\) obeys the group multiplication law: \(U(g_1)U(g_2)=U(g_1g_2)\). On the other hand, when we put the system on the open region \(M\) (\(\partial M=\{a,b\}\)), this does not necessarily hold, and we have the following relations.
\begin{equation}
    U_M(g_1)U_M(g_2)=\Omega(g_1,g_2)U_{M}(g_1g_2),
\end{equation}
where \(\Omega(g_1,g_2)\) is an operator localized at two edges of the region \(M\). By the associativity, we see that \(\Omega\) must satisfy the following conditions.
\begin{equation}
    \Omega(g_1,g_2)\Omega(g_1g_2,g_3)=\left[^{U_M(g_1)}\Omega(g_2,g_3)\right]\Omega(g_1,g_2g_3),
\end{equation}
for all \(g_1,g_2,g_3\in G\). Here, \(^xy\coloneqq xyx^{-1}\). When we restrict the \(\Omega\) to one side of the boundary, these relations hold only up to \(U(1)\) phase. Namely, there is a three-cocycle \(\omega\in C^3(G,U(1))\) such that
\begin{equation}\label{EN-3-cocycle}
    \Omega_a(g_1,g_2)\Omega_a(g_1g_2,g_3)=\omega(g_1,g_2,g_3)\left[^{U_M(g_1)}\Omega_a(g_2,g_3)\right]\Omega_a(g_1,g_2g_3),
\end{equation}
where \(\Omega_{a}\) is an operator at the edge \(a\) and comes from the restriction of \(\Omega\). Then by considering the associativity for \(\Omega\), one can show that \(\omega\in Z^3(G,U(1))\). Dividing appropriate ambiguities we can define \([\omega(g_1,g_2,g_3)]\in H^3(G,U(1))\), and this is the anomaly three-cocycle of the system.
\paragraph{Remark.}
The expression of \(\Omega\) depends on how one defines the symmetry operator for open regions, and \(\omega\) also depends on the definitions. Nevertheless, we can show that the cohomology class of \(\omega\) is independent of the choice of \(\Omega\) and the restriction \(\Omega\to\Omega_{a}\).

\subsubsection{Fermionic systems}
For fermionic systems, symmetry operators \(U(g)\) satisfies 
\begin{equation}
    U(g_1)U(g_2)=\left((-1)^F\right)^{\lambda(g_1,g_2)}U(g_1g_2),
\end{equation}
where \(\lambda\in Z^2(G,\mathbb{Z}_2)\) specifies whether the total symmetry is extended by the fermion parity symmetry. In \cite{Else:2014vma}, the authors also provided a way to extract a subclass of fermion anomalies known as Gu-Wen anomalies. Gu-Wen anomalies of \(G\) symmetries in \((1+1)\) dimensions are characterized by pairs \((\nu,n)\in C^3(G,U(1))\times H^2(G,\mathbb{Z}_2)\) such that \(\delta\nu=(-1)^{(n+\lambda)\cup n}\) (This relation is called the Gu-Wen equation). To extract these data, we again consider the symmetry action on an interval \(M=[a,b]\). Then the above multiplication law holds up to a boundary term \(\Omega\):
\begin{equation}
    U_{M}(g_1)U_{M}(g_2)=\Omega(g_1,g_2)\left((-1)^F\right)^{\lambda(g_1,g_2)}U_{M}(g_1g_2).
\end{equation}
As in the bosonic case, we again consider the restriction \(\Omega\to\Omega_{a}\).\footnote{The restricted operator \(\Omega_{a}\) is not local in the sense that fermionic operators like \(\gamma_{1},\gamma_{L}^\prime\) are not locally generated. Nevertheless, we define the restriction \(\Omega_{a}\) so that the indices of operators in \(\Omega_{a}\) are located near the point \(a\).} We define \(n\) to be \(0\) if \(\Omega_{a}\) is not charged by the fermion parity, and \(1\) if it is charged like \(\gamma_{a},\gamma_{a}^\prime\). We can show that the restricted operator \(\Omega_{a}(g_1,g_2)\) must satisfy
\begin{equation}\label{EN-3-cocycle_fermion}
    \Omega_a(g_1,g_2)\Omega_a(g_1g_2,g_3)=\nu(g_1,g_2,g_3)\left[^{U_M(g_1)}\Omega_a(g_2,g_3)\right]\Omega_a(g_1,g_2g_3),
\end{equation}
for \(\nu\in C^3(G,U(1))\). Then we can see that \(\nu\) satisfies the Gu-Wen equation \(\delta\nu=(-1)^{(n+\lambda)\cup n}\).

\section{Emergent duality and topological phase}\label{sec_emergent_dual}
\subsection{Setup and notation}
Consider a central extension of a finite group \(G\) by a finite abelian group \(A\). General such extensions fit into the short exact sequence of the form
\begin{equation}
    1 \rightarrow A \rightarrow \Gamma \rightarrow G \rightarrow 1.
\end{equation}
In particular, the isomorphic class of this extension is specified by an element of the second cohomology group \([e]\in H^2(G,A)\). We denote an element \(c\in\Gamma\) by \(c=(a,g)\). The group multiplication law is expressed as \((a_1,g_1)\cdot(a_2,g_2)=(a_1a_2e(g_1,g_2),g_1g_2)\).

To prepare the Hilbert space with this symmetry, we define the local Hilbert space \(\mathcal{H}_{j}\) as 
\begin{equation}
    \mathcal{H}_{j}=\mathbb{C}[\{\ket{c}_j\}_{c\in \Gamma}]=\mathbb{C}[\{\ket{a}_{j}\ket{g}_{j}\}_{(a,g)\in \Gamma}],
\end{equation}
where \(_{j}\braket{c|c^\prime}_{j}=\delta_{c,c^{\prime}}\)   and \(\operatorname{dim}_{\mathbb{C}}(\mathcal{H}_{j})=|\Gamma|\). We also define operator \(\widehat{c}_{j},\widehat{a}_{j},\widehat{g}_{j}\) to be 
\begin{equation}
    \widehat{c}_{j}\ket{c^\prime}_{j}=\ket{cc^\prime}_{j},\quad (\widehat{a},\widehat{g})_{j}\ket{a^\prime}_{j}\ket{g^\prime}_{j}=\ket{aa^\prime e(g,g^\prime)}_{j}\ket{gg^\prime}_{j}.
\end{equation} 
When \(\Gamma\) is abelian, we define a diagonalized operator \(c_{j}\) that satisfies \(c_{j}\ket{c^\prime}_{j}=e^{2\pi i\langle c, c^\prime\rangle}\ket{c^\prime}_{j}\). To explain the definition of \(e^{2\pi i\langle c, c^\prime\rangle}\), we assume that \(\Gamma\cong \mathbb{Z}_{n_1}\times\cdots\times\mathbb{Z}_{n_{M}}\). We denote an element \(c_i\) of \(\mathbb{Z}_{n_i}\) by \(c_i=0,\ldots,n_i-1\in\mathbb{Z}_{n_i}\). Then for \(c=(c_1,\ldots,c_M),c^\prime=(c^\prime_1,\ldots,c^\prime_M)\), we define 
\begin{equation}
    \langle c,c^\prime\rangle \coloneqq \frac{1}{n_1}c_1c_1^\prime+\cdots+\frac{1}{n_M}c_Mc_M^\prime.
\end{equation}
We see that the commutation relation between \((\widehat{c}_1)_{j}\) and \((c_2)_{j}\) is
\begin{equation}
    (\widehat{c}_1)_{j}(c_2)_{j}(\widehat{c}_1)_{j}^{-1}=e^{-2\pi i\langle c_1, c_2\rangle}(c_2)_{j}.
\end{equation}
Then we define one-dimensional Hilbert space \(\mathcal{H}\) by \(\mathcal{H}\coloneqq \bigotimes_{j=1}^{L}\mathcal{H}_{j}\). We consider an \(\Gamma\) global symmetry generated by
\begin{equation}\label{original_sym_op}
    \prod_{j=1}^{L}\widehat{c}_{j}=\prod_{j=1}^{L}\left(\widehat{a},\widehat{g}\right)_{j},
\end{equation}
and a local Hamiltonian of the form
\begin{equation}
    H_{\mathsf{D}}=\sum_{j}h_{j}.
\end{equation}
Since \(\prod_{j}\widehat{c}_{j}\) is on-site and non-anomalous, we can gauge it. In particular, we consider gauging the non-anomalous subgroup \(A\). Then the gauged Hamiltonian takes the form
\begin{equation}
    H_{\mathsf{D}/A}=\sum_{j}h_{j}^{g}
\end{equation}
with the Gauss law constraint
\begin{equation}
    (\widehat{a},\widehat{1})_{j}\ket{\psi}=a_{j-1/2}^{-1}a_{j+1/2}\ket{\psi}.
\end{equation}
Note that we have the dual fields on links and each local Hilbert space \(\mathcal{H}_{j+1/2}\) is isomorphic to \(\mathbb{C}^{|A|}\). Due to the Gauss law constraint, the \(A\) symmetry generated by \(\prod_{j=1}^{L}(\widehat{a},\widetilde{1})_{j}\) act trivially on the Hilbert space:
\begin{equation}
    \prod_{j=1}^{L}(\widehat{a},\widetilde{1})_{j}\ket{\psi}=\prod_{j=1}^{L}a_{j-1/2}^{-1}a_{j+1/2}\ket{\psi}=\ket{\psi}.
\end{equation}
Therefore, \(A\) is no longer a global symmetry but a local symmetry in the gauged theory.

To treat \(A\) as a global symmetry, we consider the following Hamiltonian:
\begin{equation}\label{eff_gauge_ham1}
    H_{\mathsf{D}/A}^K\coloneqq \sum_{j}h_{j}^g-K\sum_{j}(G_{j}+G_{j}^\dagger).
\end{equation}
Here, \(G_{j}\) is the Gauss law operator, and \(K\) is a positive constant. Note that \([h_{j}^g,G_{j^\prime}]=0,[G_{j},G_{j^\prime}]=0\) by construction. Therefore, ground states of \(H_{\mathsf{D}/A}^K\) satisfy the Gauss law constraint. Nevertheless, for a finite \(K\), the behavior \( H_{\mathsf{D}/A}^K\) as a lattice model of topological phases can be different from the model with a strict Gauss law because the symmetry operator \eqref{original_sym_op} still acts faithfully on the entire Hilbert space, i.e.~it describes a \emph{global symmetry}. The global symmetry of the Hamiltonian \(H_{\mathsf{D}/A}^K\) is \(\Gamma\times \hat{A}\) (\(\hat{A}\coloneqq \operatorname{Hom}(A,U(1))\cong A\)), and it is generated by
\begin{equation}\label{total_sym}
    U(c,b)=\prod_{j=1}^{L}\widehat{c}_{j}\prod_{j=1}^{L}\widehat{b}_{j+1/2},\quad (c,b)\in \Gamma\times \hat{A}.
\end{equation}
We refer to the procedure to obtain the Hamiltonian \eqref{eff_gauge_ham1} as \textit{effective gauging}. In the following sections, we will explore the properties of this Hamiltonian and how the \(A\) symmetry surviving under effective gauging affects the topological response action.

\subsection{Emergent anomalies}\label{sec_emergent_ano}
Since \(G_{j}\) commutes with other local interactions, one recovers the Gauss law constraint at the ground state subspace, only symmetry operators for \(G\times\hat{A}\) symmetry act faithfully on that space. We show that the system exhibits a mixed anomaly between \(G\) and \(\hat{A}\) at the ground state subspace. In particular, the anomaly three-cocycle \(e^{2\pi i\omega_0},\,\omega_0\in H^3(G\times \hat{A},U(1))\) is given by 
\begin{equation}\label{emergent_omega}
    -\omega_0\left((g_1,b_1),(g_2,b_2),(g_3,b_3)\right)=\langle b_1, \widetilde{e}(g_1,g_2)\rangle, \quad (g_i,b_i)\in G\times\hat{A}.
\end{equation}
Here we write \(\widetilde{e}(g_1,g_2)\in A\cong \hat{A}=\mathbb{Z}_{n_1}\times\cdots\times\mathbb{Z}_{n_{M(A)}}\) as 
\begin{equation}
    \widetilde{e}(g_1,g_2)=\left(e_1,\ldots,e_{M(A)}\right),
\end{equation}
so that it represents the same element as \(e(g_1,g_2)\).\footnote{For example, if \(e(g_1,g_2)=e^{2\pi ik/N}\in\mathbb{Z}_{N}\subset U(1)\), then \(\widetilde{e}(g_1,g_2)=[k]_{N}\in\mathbb{Z}_{N}=\{0,1,\ldots,N-1\}\).}

Since the Gauss law operators \(\{G_{j}\}_{j}\) commute with each local interaction \(h_{j}^g\) and they also commute with each other, the operator \((\widehat{a},\widehat{1})_j\) acts on the ground state subspace as 
\begin{equation}
    (\widehat{a},\widehat{1})_j\ket{\psi}=+a_{j-1/2}^{-1}a_{j+1/2}\ket{\psi}.
\end{equation}
Then the symmetry operators that faithfully act on the ground state subspace are generated by
\begin{equation}\label{sym_anomaly}
    U(g,b)\coloneqq \prod_{j=1}^{L}\widehat{g}_{j}\prod_{j=1}^{L}\widehat{b}_{j+1/2}.
\end{equation}
According to the Else-Nayak procedure, let us put the system on an open chain \([1/2,\ldots, L+1/2]\). We define the symmetry operators on the open chain as 
\begin{equation}
    U_{M}(g,b)\coloneqq \prod_{j=1}^{L}\widehat{g}_{j}\prod_{j=0}^{L}\widehat{b}_{j+1/2}.
\end{equation}
Then we have
\begin{align}
    \begin{split}
        U_{M}(g_1,b_1)U_{M}(g_2,b_2)&=\prod_{j=1}^{L}\left(\widehat{e}(g_1,g_2),\widehat{g_1g_2}\right)_{j}\prod_{j=0}^{L}(\widehat{b_1b_2})_{j+1/2}\\
        &=e(g_1,g_2)^{-1}_{1/2}e(g_1,g_2)_{L+1/2}U_{M}(b_1b_2,g_1g_2).
    \end{split}
\end{align}
From this, we define local operators \(\Omega_{1/2}\) as
\begin{equation}
    \Omega_{1/2}\left((g_1,b_1),(g_2,b_2)\right)=e(g_1,g_2)^{-1}_{1/2}.
\end{equation}
Then the anomaly three-cocycle \(\omega\) is calculated as 
\begin{align}
    \begin{split}
        \omega\left((g_1,b_1),(g_2,b_2),(g_3,b_3)\right)=&e(g_1,g_2)^{-1}_{1/2}e(g_1g_2,g_3)^{-1}_{1/2}e_(g_1,g_2g_3)_{1/2}\\
        &\times U_{M}(g_1,b_1)e(g_2,g_3)_{1/2}U^{-1}(g_1,b_1)\\
        =&\delta e(g_1,g_2,g_3)_{1/2}\,e^{-2\pi i \langle b_1, \widetilde{e}(g_2,g_3)\rangle}\\
        =&e^{-2\pi i \langle b_1, \widetilde{e}(g_2,g_3)\rangle},
    \end{split}
\end{align}
where we used \(\delta e=1\) in the last line. Therefore, we have found that the symmetry action \eqref{total_sym} exhibits an emergent anomaly with a three-cocycle \eqref{emergent_omega} in the ground state subspace of the Hamiltonian \eqref{eff_gauge_ham1}. Note that this is an apparent property by the construction of the Hamiltonian \eqref{eff_gauge_ham1}.

\subsubsection{Example}
Let us see an example of the mixed anomaly. We start with a non-anomalous \(\mathbb{Z}_4\) symmetry generated by
\begin{equation}
    U_{\mathbb{Z}_4}=\prod_{j=1}^{L}e^{\frac{\pi i}{4}(1-\sigma_{j}^z)}.
\end{equation}
An example of the Hamiltonian that has this symmetry is the following XY model.
\begin{equation}
    H_{\text{XY}}=-\sum_{j=1}^{L}\left(\sigma_{j}^x\sigma_{j+1}^x+\sigma_{j}^y\sigma_{j+1}^y\right).
\end{equation}
Let us gauge the non-anomalous \(\mathbb{Z}_2\) subgroup, which is generated by \(\prod_{j=1}^{L}\sigma_{j}^z\). The gauged Hamiltonian is given by
\begin{equation}
    H_{\text{XY}}^g=-\sum_{j=1}^{L}\left(\sigma_{j}^x\tau_{j+1/2}^z\sigma_{j+1}^x+\sigma_{j}^y\tau_{j+1/2}^z\sigma_{j+1}^y\right),
\end{equation}
and the Gauss law operator is \(\tau_{j-1/2}^x\sigma_{j}^z\tau_{j+1/2}^x\). By a unitary conjugation by \eqref{CZ_def} (the basis is now different), we have
\begin{equation}\label{LG_ham}
    \widetilde{H}^g_{\text{XY}}=-\sum_{j=1}^{L}\left(\tau_{j+1/2}^z-\tau_{j-1/2}^x\tau_{j+1/2}^z\tau_{j+3/2}^x\right)
\end{equation}
Then we see that the gauged system has a \(\hat{\mathbb{Z}}_2\times (\mathbb{Z}_4/\mathbb{Z}_2)\) symmetry generated by
\begin{equation}\label{z2z2_mixed}
    U_{\hat{\mathbb{Z}}_2}=\prod_{j=1}^{L}\tau_{j}^z,\quad U_{\mathbb{Z}_4/\mathbb{Z}_2}=\prod_{j=1}^{L}e^{\frac{\pi i}{4}(1-\tau_{j-1/2}^x\tau_{j+1/2}^x)}.
\end{equation}
The anomaly three-cocycle \(\omega\in H^3(\mathbb{Z}_2\times\mathbb{Z}_2,U(1))\) for this mixed anomaly is given by
\begin{equation}
    \omega((g_1,h_1),(g_2,h_2),(g_3,h_3))=(-1)^{g_1h_2h_3}.
\end{equation}
The model \eqref{LG_ham} was introduced in \cite{Levin:2012yb} as a \(\mathbb{Z}_2\) anomalous Hamiltonian and \(\mathbb{Z}_2\) symmetry is generated by the diagonal \(\mathbb{Z}_2\) subgroup of \eqref{z2z2_mixed}.

\subsection{String order and topological phase}
We are considering effective gauging and Hamiltonian of the form
\begin{equation}\label{eff_gauge_ham2}
    H_{\mathsf{D}/A}^K=\sum_{j}h_{j}^g-K\sum_{j}(G_j+G_j^\dagger).
\end{equation}
By construction, the Gauss law operators commute with \(h_{j}^g\) and act trivially on the ground state subspace. We can also change the basis by some unitary transformations so that 
\begin{equation}
    \widetilde{H}_{\mathsf{D/A}}^K\coloneqq VH_{\mathsf{D}/A}^KV^\dagger=\widetilde{H}_{\mathsf{D}}-K\sum_{j}(\widehat{g}_{j}+\widehat{g}_{j}^\dagger),
\end{equation}
where \(\widetilde{H}_{\mathsf{D}}= V\left(\sum_{j}h_{j}^g\right)V^\dagger\). Since \(V\) is a unitary operator, spectra of \(H_{\mathsf{D}/A}^K\) and \(\widetilde{H}_{\mathsf{D/A}}^K\) are the same. However, for a finite \(K\), these two models are generally in different topological phases up to the ``SPT'' phase, which is created by the entangler \(V\).

To see this, note that two ground state of \(H_{\mathsf{D}/A}^K\) and \(\widetilde{H}_{\mathsf{D}/A}^K\) satisfy
\begin{equation}
    G_{j}\ket{\text{GS}}=\ket{\text{GS}},\quad \widehat{g}_{j}\ket{\widetilde{\text{GS}}}=\ket{\widetilde{\text{GS}}},
\end{equation}
respectively. Then two ground states \(\ket{\text{GS}}\) and \(\ket{\widetilde{\text{GS}}}\) have non-zero expectation values for two string order parameters \(\prod_{j}G_{j}\) and \(\prod_{j}\widehat{g}_{j}\), i.e.,
\begin{equation}
    \bra{\text{GS}}\prod_{j\in M}G_{j}\ket{\text{GS}}\neq 0,\quad \bra{\widetilde{\text{GS}}}\prod_{j\in M}\widehat{g}_{j}\ket{\widetilde{\text{GS}}}\neq 0,
\end{equation}
for a large open region \(M\). Therefore, the ground states of two Hamiltonians \(H_{\mathsf{D}/A}^K\) and \(\widetilde{H}_{\mathsf{D}/A}^K\) belong to different topological phases with different order parameters. The string operator \(\prod_{j\in M}G_{j}\) diagnoses the nontrivial \(A\times \hat{A}\cong A\times A\) mixed SPT, known as the cluster state.
The claim here is almost trivial because the unitary transformation by \(V\), which is not an on-site symmetric operator, induces staking SPTs in general. Nevertheless, the difference is crucial when considering anomalous cases or gapless Hamiltonians. 

When \(A\) and \(G\)  form a nontrivial extension, we cannot define the \(A\times\hat{A}\) SPT phase unless the quotient \(G\) is spontaneously broken because the SPT response \(e^{-2\pi i\int (A,\hat{A})}\) carries an anomaly under the insertions of nontrivial \(G\)-defects because \(A\) is not a cocycle. In this case, the ground state subspace exhibits a mixed anomaly, which is specified by the \(3d\) SPT action \(e^{-2\pi i\int g^*e\cup\hat{A}}\).
On the other hand, \(H_{\mathsf{D}/A}^K\) is non-anomalous at the UV scale because the symmetry operator \eqref{total_sym} is on-site. Therefore, we find that we need the term \(e^{-2\pi i\int (A,\hat{A})}\) in the topological response action to describe the cancellation of the anomaly. 

When the system is gapless, we do not need to tune the value of \(K\) because the term \(\sum_{j}G_{j}\) has a gapped spectrum. In this case, the Hamiltonian \eqref{eff_gauge_ham2} describes the model for \emph{gapless SPT phases (gSPT)} \cite{Scaffidi:2017ppg,Verresen:2019igf,Thorngren:2020wet,Li:2022jbf,Wen:2022tkg} and gapless SPT phases are called \emph{intrinsically gapless SPT phases (igSPT)} if there are emergent anomalies. Note that the string order parameter \(\prod_{j\in M}G_{j}\) also works for gapless SPTs.

\subsection{Gapped phase examples}
We take the transverse-field Ising Hamiltonian \eqref{TFI_ham} as an input Hamiltonian \(H_{\mathsf{D}}\). This is the case when \(G=1,A=\hat{A}=\mathbb{Z}_2\). After effective gauging, we obtain
\begin{equation}\label{TFI_effgauge}
    H_{\text{TFI}/\mathbb{Z}_2}^K=-\sum_{j=1}^L \left(\sigma_{j}^x+J\sigma_j^z\tau_{j+1/2}^x\sigma_{j+1}^z+K\tau_{j-1/2}^z\sigma_{j}^x\tau_{j+1/2}^z\right).
\end{equation}
This model has a \(\mathbb{Z}_2^{\hat{A}}\times\mathbb{Z}_2^A\) symmetry. 
Let us discuss two cases \(J\ll 1\) and \(J\gg 1\).
\paragraph{\(J\ll 1:\)}
In this limit, \eqref{TFI_effgauge} is
\begin{equation}
    H_{\text{TFI}}^g=-\sum_{j=1}^L \left(\sigma_{j}^x+K\tau_{j-1/2}^z\sigma_{j}^x\tau_{j+1/2}^z\right).
\end{equation}
We can easily find the ground state \(\ket{\text{GS}}\). It satisfies \(\sigma_j^x\ket{\text{GS}}=+\ket{\text{GS}}, \tau_{j-1/2}^z\tau_{j+1/2}^z\ket{\text{GS}}=+\ket{\text{GS}}\) for all \(j\). Therefore, this model belongs to the \(\mathbb{Z}_2^{\hat{A}}\) SSB phase.
\paragraph{\(J\gg 1:\)}
In this limit, \eqref{TFI_effgauge} becomes
\begin{equation}
    H_{\text{TFI}}^g=-\sum_{j=1}^L \left(J\sigma_j^z\tau_{j+1/2}^x\sigma_{j+1}^z+K\tau_{j-1/2}^z\sigma_{j}^x\tau_{j+1/2}^z\right).
\end{equation}
This Hamiltonian is the cluster model \cite{Raussendorf:2001zim}, which realizes the nontrivial \(\mathbb{Z}_2^{\hat{A}}\times\mathbb{Z}_2^A\) SPT phase. When we implement the unitary transformation by \eqref{CZ_def}, we obtain
\begin{equation}
    VH_{\text{TFI}}^gV^\dagger=-\sum_{j=1}^L \left(J\tau_{j+1/2}^x+K\sigma_{j}^x\right),
\end{equation}
which belongs to the trivial phase. This reflects the fact that \eqref{CZ_def} is a \(\mathbb{Z}_2^{\hat{A}}\times\mathbb{Z}_2^A\) SPT entangler. Note that when \(K\) is infinite, the global symmetry is only \(\hat{A}\) and we do not have any SPT phases for this symmetry.

In similar manners, we can construct various cluster SPT Hamiltonians. Note that this construction is essentially the same as the decorated domain wall construction \cite{Chen:2014zvm}.

\subsection{Gapless phase examples}\label{sec_gapless_boson}
\subsubsection{TFI model}
One of the simple examples is at the critical point (\(J=1\)) in \eqref{TFI_effgauge}
\begin{equation}
    H_{\text{TFI}/\mathbb{Z}_2}^K=-\sum_{j=1}^L \left(\sigma_{j}^x+\sigma_j^z\tau_{j+1/2}^x\sigma_{j+1}^z+K\tau_{j-1/2}^z\sigma_{j}^x\tau_{j+1/2}^z\right),
\end{equation}
This gapless model describes the phase transition point between the SSB phase and the SPT phase. This model was introduced in \cite{Scaffidi:2017ppg}, see also \cite{Li:2022jbf}.

\subsubsection{XXZ chain}
We start with the XXZ Hamiltonian,
\begin{equation}
    H_{\text{XXZ}}=-\sum_{j=1}^{L}\left(\sigma_{j}^x\sigma_{j+1}^x+\sigma_{j}^y\sigma_{j+1}^y-\Delta\sigma_{j}^z\sigma_{j+1}^z\right).
\end{equation}
This model has a \(U(1)\) symmetry, and for \(-1<\Delta\leq 1\) it is described by the compact boson CFT in the IR limit. We consider the \(\mathbb{Z}_4\) subgroup generated by \(\prod_{j=1}^{L}\exp\left(\frac{\pi i}{4}(1-\sigma_{j}^z)\right)\) and gauge the \(\mathbb{Z}_2^A\) subgroup. By effective gauging, we obtain 
\begin{equation}\label{XXZ_effgauge}
    H_{\text{XXZ}/\mathbb{Z}_2}^K=-\sum_{j=1}^{L}\left(\sigma_{j}^x\tau_{j+1/2}^z\sigma_{j+1}^x+\sigma_{j}^y\tau_{j+1/2}^z\sigma_{j+1}^y-\Delta\sigma_{j}^z\sigma_{j+1}^z+K\tau_{j-1/2}^x\sigma_{j}^z\tau_{j+1/2}^x\right).
\end{equation}
The symmetry operators of emergent mixed \(\mathbb{Z}_2^{\hat{A}}\times\mathbb{Z}_2^G\) anomaly is given by \eqref{z2z2_mixed}. Therefore, this is an igSPT model. The string order parameter is given by
\begin{equation}\label{string_XXZ}
    \prod_{j=i}^{l}G_{j}=\tau_{i-1/2}^x\left(\prod_{j=i}^{l}\sigma_{i}^z\right)\tau_{l+1/2}^x.
\end{equation}
The ground state subspace of \eqref{XXZ_effgauge} is equivalent to the following Hamiltonian
\begin{equation}
    H_{\text{XXZ}}^g=-\sum_{j=1}^{L}\left(\tau_{j+1/2}^z-\tau_{j-1/2}^x\tau_{j+1/2}^z\tau_{j+3/2}^x-\Delta\tau_{j-1/2}^x\tau_{j+3/2}^x\right),
\end{equation}
and \(\mathbb{Z}_2\times\mathbb{Z}_2\) mixed anomaly symmetry operator is \eqref{z2z2_mixed}. To specify the IR CFT of this model, note that the \(\mathbb{Z}_2^A\) subgroup symmetry corresponds to the \(\mathbb{Z}_2\) shift symmetry of the compact boson. Therefore, this model is also described by a compact boson and its radius is half of the original XXZ chain. We also see that the \(\mathbb{Z}_2\times\mathbb{Z}_2\) mixed anomaly corresponds to the mixed anomaly between the \(\mathbb{Z}_2\) shift and the \(\mathbb{Z}_2\) winding symmetry of the compact boson. When \(\Delta=0\), the lattice model \eqref{XXZ_effgauge} becomes
\begin{equation}
    H_{\text{XY}}^g=-\sum_{j=1}^{L}\left(\sigma_{j}^x\tau_{j+1/2}^z\sigma_{j+1}^x+\sigma_{j}^y\tau_{j+1/2}^z\sigma_{j+1}^y+K\tau_{j-1/2}^x\sigma_{j}^z\tau_{j+1/2}^x\right),
\end{equation}
which is equivalent to the one introduced in \cite{Li:2022jbf} and the string order parameter \eqref{string_XXZ} is equivalent to what introduced in \cite{Wen:2022tkg,Wen:2023otf}.

\subsubsection{Clock model}
We consider the following \(n\)-level spin clock model (\(n\geq 2\)):
\begin{equation}\label{clock_ham}
    H_{\text{clock}}=-\sum_{j=1}^{L}\left(X_{j}+JZ_{j}^\dagger Z_{j+1}\right)+\mathrm{h.c.},
\end{equation}
where \(X_{j}\) and \(Z_{j}\) satisfy relations: \(X_{j}^n=Z_{j}^n=I_{n},Z_{j}X_{j}=e^{2\pi i/n}X_{j}Z_{j}\). \(I_n\) is the \(n\times n\) identity matrix. It is known that, for \(2\leq n\leq 4\), \(J=1\) is a critical point, while for larger \(n\) the model is gapless over a finite range of \(J\). When \(n\geq 5\), the underlying IR theory is the \(U(1)_{2n}\) Wess-Zumino-Witten CFT, which is realized as a compact boson theory at a specific radius \cite{Li:2014pta}.
The model has a \(\mathbb{Z}_n\) global symmetry generated by \(\prod_{j}X_{j}\). Let us effective gauge the \(\mathbb{Z}_{m}\) subgroup (suppose \((m\mid n)\)). We obtain
\begin{equation}
    H_{\text{clock}/\mathbb{Z}_m}^K=-\sum_{j=1}^{L}\left(X_{j}+JZ_{j}^\dagger \widetilde{X}_{j+1/2} Z_{j+1}+K\widetilde{Z}_{j-1/2}X_{j}^{n/m}\widetilde{Z}_{j+1/2}^\dagger\right)+\mathrm{h.c.},
\end{equation}
where \(\widetilde{X}_{j+1/2},\widetilde{Z}_{j+1/2}\) denote the \(m\)-level spin operators. Whether this model is an igSPT or not depends on the choice of \(n\) and \(m\). For example when \(n=4,m=2\), this is an igSPT model and exhibits deconfined quantum criticality in the IR. The IR model was studied in \cite{Zhang:2022wwn,Su:2023hud}.

\subsubsection*{Comments}
\begin{itemize}
    \item When the original system with a Hamiltonian \(H_{\mathsf{D}}\) is gapless, the dual system with \(H_{\mathsf{D}/A}\) or \(H_{\mathsf{D}/A}^K\) is also gapless. To see this, we note that the gaugings of finite symmetries are done by considering insertions of all topologically distinct classes of topological defects. Such insertions just twist the boundary conditions of the systems, and we expect gapless spectrums to be stable against such boundary conditions.
    \item Though we assumed the original symmetry \(\Gamma\) is non-anomalous (on-site) so far, this assumption is not essential and we only need \(A\) to be non-anomalous. We will consider such an anomalous case in the next section.
    \item In our construction \eqref{eff_gauge_ham2}, all emergent anomalies are mixed type and not all mixed anomalies appear from gauging subgroups of non-anomalous theories. Nevertheless, by considering subgroups of \(\hat{A}\times G\), we obtain various other anomalies. For example, when we take the diagonal subgroup \(\mathbb{Z}_2^{\hat{A}}\times \mathbb{Z}_2^G\) in \eqref{XXZ_effgauge}, then we obtain a pure \(\mathbb{Z}_2\) anomaly.
\end{itemize}

\section{Fermionic theory}\label{sec_fermion}
\subsection{Non-anomalous gapped phase}\label{sec_non-ano_fermion}
We can implement \textit{effective fermionization} in the same split as effective gauging. We consider a \(\Gamma\) global symmetry and \(\Gamma\) is extended by non-anomalous \(\mathbb{Z}_2^A\) subgroup as follows.
\begin{equation}
    1 \rightarrow \mathbb{Z}_2^A \rightarrow \Gamma \rightarrow G \rightarrow 1.
\end{equation}
After fermionization, the system has the fermion parity \(\mathbb{Z}_2^F\) symmetry and the \(G\) symmetry. We define \emph{effective fermionization} by the following Hamiltonian:
\begin{equation}\label{fer_effgauge}
    H_{\mathcal{F}(\mathsf{D})}^K=\sum_{j=1}^{L}h^f_{j}-K\sum_{j=1}^{L}G_{j}^f.
\end{equation}
This Hamiltonian has the \(\mathbb{Z}_2^A\) symmetry for a finite \(K\).

We first discuss the case when \(G\) is trivial. We take the TFI Hamiltonian \eqref{TFI_ham} as an input. Then \eqref{fer_effgauge} becomes
\begin{equation}\label{TFI_Feffgauge}
    H_{\mathcal{F}(\text{TFI})}^K=-\sum_{j=1}^L \left(\sigma_{j}^x+J\sigma_j^z(-i\gamma_{j+1/2}\gamma_{j+1/2}^\prime)\sigma_{j+1}^z+K(-i\gamma_{j-1/2}^\prime\gamma_{j+1/2})\sigma_j^x\right).
\end{equation}
Let us examine two cases \(J\ll 1\) and \(J\gg 1\).
\paragraph{\(J\ll 1:\)} The Hamiltonian \eqref{TFI_Feffgauge} goes to
\begin{equation}
    H_{\mathcal{F}(\text{TFI})}^K=-\sum_{j=1}^L \left(\sigma_{j}^x+K(-i\gamma_{j-1/2}^\prime\gamma_{j+1/2})\sigma_j^x\right).
\end{equation}
The ground state \(\ket{\text{GS}}\) of this Hamiltonian satisfies \(\sigma_{j}^x\ket{\text{GS}}=\ket{\text{GS}}\) and \(-i\gamma_{j-1/2}\gamma_{j+1/2}^\prime\ket{\text{GS}}=\ket{\text{GS}}\). Therefore, \(\ket{\text{GS}}\) belongs to (\(\mathbb{Z}_2^A\) trivial)\(\times\)(Kitaev phase). This phase is unchanged under the transformation by \(V_f\) \eqref{Vf}. Indeed, the Hamiltonian is mapped as
\begin{equation}
    V_{f}H_{\mathcal{F}(\text{TFI})}^KV_{f}^\dagger=-\sum_{j=1}^{L}\left((-i\gamma_{j-1/2}^\prime\gamma_{j+1/2})\sigma_j^x+K\sigma_{j}^x\right),
\end{equation}
and \(V_{f}\ket{\text{GS}}\) also satisfies the same relations as \(\ket{\text{GS}}\).

\paragraph{\(J\gg 1:\)} The Hamiltonian \eqref{TFI_Feffgauge} goes to
\begin{equation}
    H_{\mathcal{F}(\text{TFI})}^K=\sum_{j=1}^L \left(J\sigma_j^z(-i\gamma_{j+1/2}\gamma_{j+1/2}^\prime)\sigma_{j+1}^z+K(-i\gamma_{j-1/2}^\prime\gamma_{j+1/2})\sigma_j^x\right).
\end{equation}
This Hamiltonian describes the nontrivial \(\mathbb{Z}_2^A\times\mathbb{Z}_2^F\) SPT phase. Unitary conjugation by \(V_f\) maps the model to
\begin{equation}
    V_fH_{\mathcal{F}(\text{TFI})}^KV_f^\dagger=-\sum_{j=1}^{L}\left(J(-i\gamma_{j+1/2}\gamma_{j+1/2}^\prime)+K\sigma_{j}^x\right),
\end{equation}
which describes the trivial \(\mathbb{Z}_2^A\times\mathbb{Z}_2^F\) phase.

Comparing the above two cases, the interpretation of ``conjugation by \(V_f\)'' is unclear and depends on the choice of the original Hamiltonian \(H_{\mathsf{D}}\). In the example of \(H_{\text{TFI}}\), \(V_{f}\) changes \(Z_{\mathcal{F}(\text{TFI})}[\rho,A]=(-1)^{\operatorname{Arf}(\rho)}\mapsto(-1)^{\operatorname{Arf}(\rho)}\) when \(J\ll 1\), while \(V_{f}\) maps \((-1)^{\operatorname{Arf}(A+\rho)+\operatorname{Arf}(\rho)}\mapsto 1\) when \(J\gg 1\).
The reason for this will be clarified in the field theory formalism, which we discuss in Sec.~\ref{sec_fermionQFT}.  Note that the behavior under the strict Gauss law is determined by fermionization procedures and is always well-defined.

\subsection{Fermionizing \texorpdfstring{\(\mathbb{Z}_2\)}{Z2} subgroups}
One may expect that fermionizing a non-anomalous \(\mathbb{Z}_2\) subgroup with a nontrivial extension \(1\rightarrow \mathbb{Z}_2\rightarrow \Gamma\rightarrow G\rightarrow 1\) induces a fermionic anomaly with \(G\) symmetry. However, in this case, the fermionized theory does not have any anomalies, and \(G\) global symmetry in the theory is extended by the fermion parity symmetry \cite{Inamura:2022lun}.\footnote{Though the phase \(z(\rho,a;b)\) or \(z(\rho,a)\) (in \eqref{usual_fer_qft} and \eqref{arf_intro}) seems to carry an anomaly, the bulk dependence is actually trivialized, see Sec.~\ref{sec_fer_subgp} for this point.} Here, we explain this for \(\Gamma=\mathbb{Z}_{2n}\) symmetries.

Suppose that we have on-site operators \(\{X_j\}_{j}\) such that \(X_{j}^{2n}=1\). Consider a \(\mathbb{Z}_{2n}\) symmetry generated by \(U_{\mathbb{Z}_{2n}}=\prod_{j=1}^{L}X_j\). Then we fermionize the \(\mathbb{Z}_2\) subgroup symmetry generated by \(U_{\mathbb{Z}_{2n}}^n\). The Gauss law is
\begin{equation}
    X_{j}^n(-i\gamma_{j-1/2}^\prime\gamma_{j+1/2})\ket{\psi}=+\ket{\psi},
\end{equation}
and from this, we see that the symmetry operator \(U\) satisfies
\begin{equation}
    U_{\mathbb{Z}_{2n}}^n=\prod_{j=1}^{L}X_{j}^n=\prod_{j=1}^{L}\left(-i\gamma_{j-1/2}^\prime\gamma_{j+1/2}\right)=(-1)^{t_f}(-1)^F
\end{equation}
in the fermionized theory. Here \(t_f=0,1\) specifies the boundary condition of Majorana operators so that \(\gamma_{L+1/2}=-(-1)^{t_f}\gamma_{1/2}\). Therefore, the \(\mathbb{Z}_n\) symmetry is extended by the fermion parity symmetry in the fermionized theory.

\subsection{\texorpdfstring{\(\mathbb{Z}_2\)}{Z2} fermion anomaly}\label{sec_Z2fermionanomaly}
To obtain \(G\times\mathbb{Z}_2^F\) anomalous theories by fermionizing some \(\mathbb{Z}_2\) subgroup, we need to prepare anomalous systems. Let \([e]\neq 0\) be an element of \(H^2(BG,\mathbb{Z}_2)\) that specifies the extension. We use anomaly three-cocycles with the \(\Gamma\) symmetry of the form
\begin{equation}
    \omega=\frac{1}{2}A\cup g^*e+g^*\nu.
\end{equation}
Here, \(\nu\in C^3(BG,\mathbb{R}/\mathbb{Z})\) satisfies \(\delta\nu=\frac{1}{2}e\cup e\), which ensures \(\omega\) is a cocycle. We detail fermionizing such anomalies in Sec.~\ref{sec_QFT}. Here, we discuss the case when \(G=\mathbb{Z}_2\), and take \(g^*e=G\cup G\) for \(\mathbb{Z}_2\) gauge field \(G\).

The \(\Gamma\cong\mathbb{Z}_4\) symmetry operator that describes such an anomaly is given by
\begin{equation}
    U=\prod_{j=1}^{L}\widehat{g}_{j}\prod_{j=1}^{L}S_{j,j+1}.
\end{equation}
The first term \(\prod_{j}\widehat{g}_{j}\) describes an on-site \(\mathbb{Z}_4\) symmetry and the operator \(S_{j,j+1}\) is defined as
\begin{equation}
    S_{j,j+1}=e^{2\pi i\left([g_{j}-g_{j+1}]_{4}\right)/8},
\end{equation}
where \(g_{j}\) is a diagonalized operator and its eigenvalues are in \(\{0,\ldots,3\}\). \([\cdot]_4\) means we evaluate modulo 4. In the following, we construct \(\mathbb{Z}_2\times\mathbb{Z}_2^F\) fermionic symmetry operators with the Gu-Wen fermionic anomaly. To construct such operators, we gauge the non-anomalous \(\mathbb{Z}_2\) symmetry. After fermionization, the operator \(\widehat{g}_{2}^2\) is identified as 
\begin{equation}
    \widehat{g}_{j}^2=-i\gamma_{j-1/2}^\prime\gamma_{j+1/2}.
\end{equation}
Since the operator \(S_{j,j+1}\) is charged by \(\widehat{g}_{j}^2\), it is modified under fermionization as 
\begin{equation}
    \widetilde{S}_{j,j+1}=e^{2\pi i\left([g_{j}-2n_{j+1/2}-g_{j+1}]_{4}\right)/8},
\end{equation}
where \(n_{j+1/2}\in\{0,1\}\) specifies the fermion parity at \(j+1/2\). Then we obtain the \(\mathbb{Z}_2\) symmetry operator \(U_{\mathsf{F}}\) in the fermionized theory as
\begin{equation}
    U_{\mathsf{F}}=i^{t_f}\prod_{j=1}^{L}\widehat{g}_{j}\prod_{j=1}^{L}\widetilde{S}_{j,j+1/2,j+1}.
\end{equation}
As we see in the following, this symmetry action realizes the Gu-Wen \(\mathbb{Z}_2\) fermionic anomaly, which is specified by a pair \(\nu(g_1,g_2,g_3)=i^{g_1g_2g_3},n(g_1,g_2)=(-1)^{g_1g_2}\). To diagnose the anomaly, we use the Else-Nayak procedure for fermionic systems. Let us put the system on an open chain \(\{1/2,\ldots,L+1/2\}\) and define the boundary symmetry operator as
\begin{equation}
    U_{\mathsf{F},M}\coloneqq i^{t_f}\prod_{j=1}^{L}\widehat{g}_{j}\prod_{j=1}^{L-1}\widetilde{S}_{j,j+1/2,j+1}.
\end{equation}
Then we have
\begin{align}
    \begin{split}
        U_{\mathsf{F},M}^2&=(-1)^{t_f}\prod_{j=1}^{L}\left(-i\gamma_{j-1/2}^\prime\gamma_{j+1/2}\right)\prod_{j=1}^{L-1}e^{\frac{2\pi i}{4}[g_{j}-2n_{j+1/2}-g_{j+1}]_4}\\
        &=-(-1)^{t_f}(-i\gamma_{1/2}\gamma_{j+1/L}^\prime)(-1)^Fe^{\frac{2\pi i}{4}g_{1}-\frac{2\pi i}{4}g_{L+1/2}}(-i\gamma_{1/2}\gamma_{1/2}^\prime)(-1)^{F}(-i\gamma_{L+1/2}\gamma_{L+1/2}^\prime)\\
        &=-i(-1)^{t_f}\gamma_{1/2}^\prime\,e^{\frac{2\pi i}{4}g_{1}}\gamma_{L+1/2}\,e^{-\frac{2\pi i}{4}g_{L+1/2}}.
    \end{split}
\end{align}
According to the Else-Nayak procedure, we define local operator \(\Omega_{1/2}(g,h)\,(g,h\in\{0,1\})\) as
\begin{equation}
    \Omega(g,h)_{1/2}\coloneqq
    \begin{dcases}
        \gamma_{1/2}^\prime\,e^{\frac{2\pi i}{4}g_{1}} & g=h=1,\\
        1& \text{other}.
    \end{dcases}
\end{equation}
From them, we find that \(n(g_1,g_2)=(-1)^{g_1g_2}\), and from \eqref{EN-3-cocycle_fermion}, we obtain
\begin{align}
    \begin{split}
        \nu(1,1,1)&=\Omega_{1/2}(1,1)\left(U_{\mathsf{F},M}\Omega_{1/2}(1,1)^{-1}U_{\mathsf{F},M}^{-1}\right)\\
        &=i,
    \end{split}
\end{align}
and the others are trivial. The pair \((\nu,n)\) realizes the nontrivial \(\mathbb{Z}_2\) Gu-Wen fermionic anomaly.

\subsection{Fermionic gapless phase}\label{sec_fermion_gapless}
We introduce fermionic gapless SPT model examples. An example of fermionic gapless SPT models was studied in \cite{Borla:2020avq}. In this section, we introduce two examples, without an emergent anomaly model and with the emergent Gu-Wen fermionic anomaly.

\subsubsection{No emergent anomaly example}
This is given by the TFI Hamiltonian as an input \(H_{\mathsf{D}}\). The model is the critical point of \eqref{TFI_Feffgauge}. 
\begin{equation}
    H_{\mathcal{F}(\text{TFI})}^K=-\sum_{j=1}^L \left(\sigma_{j}^x+\sigma_j^z(-i\gamma_{j+1/2}\gamma_{j+1/2}^\prime)\sigma_{j+1}^z+K(-i\gamma_{j-1/2}^\prime\gamma_{j+1/2})\sigma_j^x\right).
\end{equation}
Its low energy spectrum is equal to the critical point of \eqref{fer_TFI_ham}:
\begin{equation}
    \widetilde{H^\prime}_{\mathcal{F}(\text{TFI})}=\sum_{j=1}^{L}\left(i\gamma_{j-1/2}^\prime\gamma_{j+1/2}+i\gamma_{j+1/2}\gamma_{j+1/2}^\prime\right),
\end{equation}
which is described by the fermionization of the Ising CFT, i.e.~free Majorana CFT in the IR.

\subsubsection{Emergent anomaly example}
Similar to bosonic cases, we have fermionic gapless SPT models with emergent anomalies and an example of such models was given in \cite{Borla:2020avq}. Here, we give an example of fermionic intrinsically gapless SPT models with a \(\mathbb{Z}_2\) Gu-Wen anomaly. As explained in Sec.~\ref{sec_Z2fermionanomaly}, the bosonized system of the Gu-Wen fermion anomaly is also anomalous. On the other hand, one of the features of gapless SPTs is that global symmetries are non-anomalous at the UV scale. To overcome the difficulty, we start with a non-anomalous \(\mathbb{Z}_8\) symmetry. We summarize our strategy as follows:
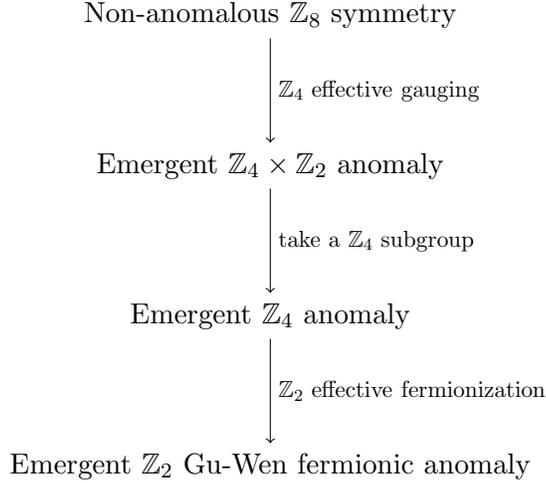
\begin{equation*}
    \begin{tikzpicture}
        \centering
        \pgfmathsetmacro{\sepy}{2}
        \node[] (1) at (0,0) {Non-anomalous \(\mathbb{Z}_8\) symmetry};
        \node[] (2) at (0,-\sepy) {Emergent \(\mathbb{Z}_4\times\mathbb{Z}_2\) anomaly};
        \node[] (3) at (0,-2*\sepy) {Emergent \(\mathbb{Z}_4\) anomaly};
        \node[] (4) at (0,-3*\sepy) {Emergent \(\mathbb{Z}_2\) Gu-Wen fermionic anomaly};
        \draw[->] (1) -- node[scale=.8,anchor=west]{\(\mathbb{Z}_4\) effective gauging} (2);
        \draw[->] (2) -- node[scale=.8,anchor=west]{take a \(\mathbb{Z}_4\) subgroup} (3);
        \draw[->] (3) -- node[scale=.8,anchor=west]{\(\mathbb{Z}_2\) effective fermionization} (4);
    \end{tikzpicture}
\end{equation*}
To construct the model, we use the \(\mathbb{Z}_8\) clock model. The global symmetry is generated by
\begin{equation}
    U=\prod_{j=1}^{L}X_{j},
\end{equation}
where \(X_{j}\) is a eight-level spin at \(j\). The gapless Hamiltonian with this symmetry is
\begin{equation}
    H_{\text{clock}}=-\sum_{j=1}^{L}\left(X_{j}+JZ_{j}^\dagger Z_{j+1}\right)+\mathrm{h.c.},
\end{equation}
see \eqref{clock_ham} for the definition of each operator. We gauge the non-anomalous \(\mathbb{Z}_4\) subgroup generated by \(U^2\) and use four-level spins for dual degrees. The effective gauged Hamiltonian is given by
\begin{equation}
    H_{\text{clock}/\mathbb{Z}_4}^{K}=-\sum_{j=1}^{L}\left(X_{j}+JZ_{j}^\dagger\widetilde{X}_{j+1/2}Z_{j+1}+K\widetilde{Z}_{j-1/2} X_{j}^2 \widetilde{Z}_{j+1/2}^\dagger\right)+\mathrm{h.c.}
\end{equation}
This model has a \(\mathbb{Z}_8\times\mathbb{Z}_4\) symmetry generated by
\begin{equation}
    U=\prod_{j=1}^{L}X_{j},\quad \widetilde{U}=\prod_{j=1}^{L}\widetilde{X}_{j+1/2}. 
\end{equation}
These symmetry operators carry a \(\mathbb{Z}_4\times\mathbb{Z}_2\) emergent anomaly as we saw in Sec.~\ref{sec_emergent_ano}. To go to the next step, we change the basis by a unitary conjugation. Specifically, we use the following unitary operator \(V\).
\begin{equation}
    V=\prod_{j=1}^{L}e^{\frac{2\pi i}{8}g_{j}[-\widetilde{h}_{j-1/2}+\widetilde{h}_{j+1/2}]_{4}},
\end{equation}
where \(g_{j}\in\{0,\ldots,7\}\,(\widetilde{h}_{j+1/2}\in\{0,\ldots,3\})\) is a value of eight(four)-level spin in the \(Z(\widetilde{Z})\) basis. Then we obtain
\begin{equation}
    \widetilde{H}_{\text{clock}/\mathbb{Z}_4}^{K}=V H_{\text{clock}/\mathbb{Z}_4}^{K}V^\dagger=-\sum_{j=1}^{L}\left(X_{j}\widetilde{S}_{j-1/2,j+1/2}+JY_{j}\widetilde{X}_{j+1/2}Y_{j+1}^\dagger+KX^2_{j}\right)+\mathrm{h.c.}
\end{equation}
The operator \(\widetilde{S}_{j-1/2,j+1/2}\) is \(\widetilde{S}_{j-1/2,j+1/2}=e^{\frac{2\pi i}{8}[-\widetilde{h}_{j-1/2}+\widetilde{h}_{j+1/2}]_{4}}\). The operator \(Y_{j}\) is complicated but only depends on \(g_{j}\) and \([-\widetilde{h}_{j-1/2}+\widetilde{h}_{j+1/2}]_4\). 
The unitary conjugation by \(V\) also changes the symmetry operator \(U\). 

Then we take a emergent anomalous \(\mathbb{Z}_4\) subgroup of the \(\mathbb{Z}_4\times\mathbb{Z}_2\) symmetry. The \(\mathbb{Z}_4\) symmetry is generated by 
\begin{equation}\label{Z4_condense}
    V\widetilde{U}UV^\dagger=\prod_{j=1}^{L}\widetilde{X}_{j+1/2}\prod_{j=1}^{L}X_{j}\widetilde{S}_{j-1/2,j+1/2},
\end{equation}
which gives the emergent anomalous \(\mathbb{Z}_4\) symmetry we saw in Sec.~\ref{sec_Z2fermionanomaly}. Note that for a finite \(K\), the symmetry is non-anomalous though \(V\widetilde{U}UV^\dagger\) is not an on-site operator. To obtain an igSPT model with the \(\mathbb{Z}_2\) Gu-Wen fermionic anomaly, we fermionize the non-anomalous \(\mathbb{Z}_2\) symmetry generated by \(\prod_{j}\widetilde{X}_{j+1/2}^2=\left(V\widetilde{U}UV^\dagger\right)^2\). By effective fermionization, we obtain 
\begin{equation}
    VH_{\text{figSPT}}V^\dagger=-\sum_{j=1}^{L}\left(X_{j}\widetilde{S}_{j-1/2,j,j+1/2}+J\widetilde{Y}_{j}\widetilde{X}_{j+1/2}\widetilde{Y}_{j+1}^\dagger+KX^2_{j}+K_{f}\widetilde{X}_{j+1/2}^2(-i\gamma_{j}^\prime\gamma_{j+1})\right)+\text{h.c.},
\end{equation}
where \(\widetilde{S}_{j-1/2,j,j+1/2}\) and \(\widetilde{Y}_{j}\) are modified operators so that they commute with the Gauss law operators \(\{\widetilde{X}_{j+1/2}^2(-i\gamma_{j}^\prime\gamma_{j+1})\}_{j}\). The symmetry operator for the emergent anomaly is given by
\begin{equation}
    i^{t_f}\prod_{j=1}^{L}\widetilde{X}_{j+1/2}\prod_{j=1}^{L}X_{j}\widetilde{S}_{j-1/2,j,j+1/2}.
\end{equation}
On the original basis, \(H_{\text{figSPT}}\) is written as 
\begin{align}
    \begin{split}
        H_{\text{figSPT}}=-\sum_{j=1}^{L}&\left(X_{j}\widetilde{S}_{j-1/2,j+1/2}^{\,\dagger}\widetilde{S}_{j-1/2,j,j+1/2}+J\widetilde{Y}_{j}Y_{j}^\dagger\widetilde{X}_{j+1/2}Y_{j+1}\widetilde{Y}_{j+1}^\dagger+K\widetilde{Z}_{j-1/2}X^2_{j}\widetilde{Z}_{j+1/2}^\dagger\right.\\
        &+\left.K_{f}(Y_{j}^\dagger)^2\widetilde{X}_{j+1/2}^2(Y_{j+1})^2(-i\gamma_{j}^\prime\gamma_{j+1})\right)+\text{h.c.}
    \end{split}
\end{align}
In summary, the fermionic igSPT Hamiltonian \(H_{\text{figSPT}}\) has a \(\mathbb{Z}_8\times\mathbb{Z}_4\times\mathbb{Z}_2^F\) global symmetry at the UV scale and exhibits the emergent \(\mathbb{Z}_2\) Gu-Wen fermionic anomaly in the IR.

\section{Higher dimensions}\label{sec_highdim}
Our construction can be immediately generalized to higher dimensions. If we start with a global zero-form symmetry in spacetime \(d\) dimensions, the emergent anomaly is a mixed anomaly between zero-from and \((d-2)\)-form symmetries. A simple example of such lattice models in \((2+1)\) dimensions was discussed in \cite{Verresen:2022mcr}. The little difference to our construction is they imposed the Gauss law strictly on the state. When we introduce the Gauss law as an energy cost, we obtain a lattice model for a mixed \(\mathbb{Z}_2\) one-form and \(\mathbb{Z}_2\) zero-form SPT phase. We can also construct emergent anomalous models from nontrivial symmetry extensions.

In higher dimensions, we can also consider emergent higher-group symmetries. For example, consider a \(G_1\times G_2\) mixed anomaly in a \((2+1)\)-dimensional system. Suppose the anomaly is \(\int_{X_{4}} G_1\cup \beta(G_2)\). If we implement effective gauging for \(G_1\), we obtain a model for emergent two-group symmetry and \(\beta\) becomes the Postnikov class. If one wants to start from more general zero-form anomalies, one can use lattice models with the anomalies, as detailed in Appendix \ref{sec_construct_ano_ham}.

We can also define fermionization and bosonization in higher dimensions. See \cite{Kapustin:2017jrc,Thorngren:2018bhj} for field theory formalism and \cite{Chen:2017fvr,Chen:2018nog,Chen:2019wlx} for lattice systems. It would be interesting to apply such fermionization schemes to our constructions in this paper.

\section{Field theory perspective}\label{sec_QFT}
\subsection{Gauging non-anomalous subgroups}
We start with very simple cases. We will discuss more general cases later. Consider a spacetime \(d\)-dimensional system with a non-anomalous zero-form global finite \(\Gamma\) symmetry, which obeys the following cental extension. 
\begin{equation}\label{extension_qft}
    1 \rightarrow A \rightarrow \Gamma \rightarrow G \rightarrow 1.
\end{equation}
We will denote the extension class by \([e] \in H^2(BG,A)\).

When the extension is nontrivial, the cocycle condition of the background field \(A\in C^1(X_d,A)\) for the \(A\) symmetry%
\footnote{In this section, we use the same symbols for symmetry groups and the corresponding gauge fields. Note that, in this notation, the trivial group \(G=1\) has the trivial gauge field \(G=0\).}
should be modified:
\begin{equation}
    \delta A=-g^*e,
\end{equation}
where \(g\) is a classifying map. Due to this condition, the gauged theory exhibits an 't Hooft anomaly even though there is no anomaly in the original theory. The anomaly of the gauged theory is specified by the \((d+1)\)-dimensional bulk action of the form \cite{Tachikawa:2017gyf}
\begin{equation}
    -\int_{X_{d+1}}g^*e\cup\hat{A},
\end{equation}
where \(\hat{A}\in H^{d-1}(X_d,\hat{A})\) is a background gauge field for the dual symmetry \(\hat{A}=\operatorname{Hom}(A,U(1))\cong\hat{A}\).

\paragraph{Example.}
Consider \((1+1)\)-dimensional theories and take \(G = A=\mathbb{Z}_2\). The nontrivial choice of \([e]\in H^2(B\mathbb{Z}_2,\mathbb{Z}_2)\cong\mathbb{Z}_2\) is \(g^*e=G\cup G\), where \(\mathbb{Z}_2\) gauge field \(G\) is normalized so that \(\oint G=\{0,1\}\). When we gauge the \(\mathbb{Z}_2^{A}\) symmetry, the gauged theory has an anomaly of the form
\begin{equation}
    -\frac{1}{2}\int_{X_3}G\cup G\cup\hat{A}.
\end{equation}

\subsection{Our construction}
We start with a theory \(\mathsf{D}\) with a non-anomalous zero-form \(\Gamma\) global symmetry. Here, \(\Gamma\) obeys the central extension \eqref{extension_qft}, and the data of the extension is specified by the relation \(\delta A=-g^*e\) for some \([e]\in H^{2}(BG,A)\). In the absence of \(G\)-defects, the gauged theory has a topological operator \(\exp\left(\oint_{X_{p+1}}a\right)\) and the partition function is given by
\begin{equation}
    Z_{\mathsf{D}/A}[\hat{A},G=0;e]=\#\sum_{a}Z_{\mathsf{D}}[G=0,a;e]\,e^{2\pi i\int_{X_d} (a,\hat{A})}.
\end{equation}
In the presence of \(G\)-defects, the partition function is no longer invariant under the shift of the dual field \(\hat{A}\) and the anomaly is captured by the bulk SPT action \(-2\pi i\int_{X_{d+1}} g^*e\cup A\).

We consider the Hamiltonian of the form
\begin{equation}
    H_{\mathsf{D}/A}^K=\sum_{j}h_j^g-K\sum_{j}(G_j+G_j^\dagger).
\end{equation}
See Sec.~\ref{sec_intro} and Sec.~\ref{sec_emergent_dual} for the definition. Our claim is that this Hamiltonian describes the theory of 
\begin{equation}\label{effgauge_Z}
    Z_{\mathsf{D}/A}[\hat{A},G,A;e]=\#\sum_{a}Z_{\mathsf{D}}[G,a;e]\,e^{2\pi i\int_{X_d} (a-A,\hat{A})}=Z_{\mathsf{D}/A}[\hat{A},G;e]\,e^{-2\pi i\int_{X_{d}}(A,\hat{A})}.
\end{equation}
Clearly, the anomaly in \(Z_{\mathsf{D}/A}[\hat{A},G;e]\) is cancelled by the additional term \(\exp\left(-2\pi i\int_{X_{d}}(A,\hat{A})\right)\) and the story is consistent with the fact that \(H_{\mathsf{D}/A}^K\) does not have any anomalies at the UV scale. Another understanding is that two gauge fields \(a\) and \(A\) should couple to the dual field \(\hat{A}\) almost in the same way because symmetry operators for \(a\) and \(A\) symmetry are realized by the same operator on the lattice. 

Note that the global symmetry \(A\) acts trivially on the IR theory because the additional term is almost decoupled to the gauged theory \(Z_{\mathsf{D}/A}[\hat{A},G;e]\), and the system exhibits the emergent anomaly as expected.

When the \(G\) symmetry is spontaneously broken, the term \(\exp\left(-2\pi i\int_{X_{d}}(A,\hat{A})\right)\) just describes an SPT phase. When \(\hat{A}\) is broken, the theory becomes just some \(G\) symmetric theory. The remained interesting class is that both \(G\) and \(\hat{A}\) are unbroken in the IR.

\subsection{Gapless phase}
Consider the theory of the form \eqref{effgauge_Z}. Suppose that the original theory \(\mathsf{D}\) is gapless and in particular described by some conformal field theory (CFT). Then the gauged theory \(\mathsf{D}/A\) is also described by CFT. In this case, \eqref{effgauge_Z} is called \emph{gapless SPT phase (gSPT)}. In particular, gSPT is called \textit{intrinsically gapless SPT (igSPT)} \cite{Thorngren:2020wet} when \([e]\) is nontrivial. In general, a \(\Gamma\) symmetric gapless SPT model is characterized by a symmetry group extension 
\begin{equation}
    1\rightarrow G_{\text{gap}} \rightarrow \widetilde{\Gamma} \rightarrow G_{\text{low}} \rightarrow 1,
\end{equation}
where the \(G_{\text{gap}}\) symmetry is decoupled in the IR. In our construction \eqref{effgauge_Z}, \(G_{\text{gap}}=A\), \(G_{\text{low}}=\hat{A}\times G\), and \(\widetilde{\Gamma}=\hat{A}\times\Gamma\), and the emergent anomalies are always mixed anomalies. Nevertheless, we can construct pure anomalous cases by condensations, i.e.~taking appropriate subgroups of \(\hat{A}\times G\) and \(\hat{A}\times\Gamma\).

\subsubsection{Relation to Li-Oshikawa-Zheng \cite{Li:2023knf}}
In \cite{Li:2023knf}, the authors pointed out that the \(\mathbb{Z}_4\) igSPT model is constructed by the combination of some duality transformations. Let us see how their construction relates to our formalism. Suppose that a \((1+1)\)-dimensional theory \(\mathcal{X}\) has a global \(A\times B\) zero-form symmetry. We assume that both \(A\) and \(B\) are finite abelian groups and they are isomorphic. Let us define two topological manipulations \(S\) and \(T\) as follows.
\begin{gather}
    Z_{S\mathcal{X}}[A,B]\coloneqq \#\sum_{a,b}Z_{\mathcal{X}}[a,b]\,e^{2\pi i\int (a,B)+(b,A)},\\
    Z_{T\mathcal{X}}[A,B]\coloneqq Z_{\mathcal{X}}[A,B]\,e^{2\pi i\int (A,B)}.
\end{gather}
We start with a \(\Gamma\times B\) global symmetry, where \(\Gamma\) is specified by the extension class \([e]\in H^2(G,A)\). As an input, we prepare the following theory.
\begin{equation}
    Z_{\mathsf{D}}[G,A,B]\coloneqq Z_{\mathcal{G}}[G,A]\times\delta(B),
\end{equation}
where \(\mathcal{G}\) is a gapless theory. We first implement the \(S\) operation to this theory. Then we have
\begin{equation}
    Z_{S\mathsf{D}}[G,A,B]=Z_{\mathcal{G}/A}[G,B]\times 1.
\end{equation}
Note that \(\mathcal{G}/A\) has a mixed anomaly between \(G\) and \(B\). Next, we implement \(ST^{-1}\) operation to \(S\mathsf{D}\) and obtain
\begin{align}
\begin{split}
    Z_{ST^{-1}(S\mathsf{D})}[G,A,B]
    &=\#\sum_{a,b}Z_{\mathcal{G}/A}[G,b]\,e^{-2\pi i\int(a,b)}e^{2\pi i\int(a,B)+(b,A)}\\
    &=\#\sum_{b}Z_{\mathcal{G}/A}[G,b]\,\delta(-b+B)\,e^{2\pi i\int(b,A)}\\
    &=\#Z_{\mathcal{G}/A}[G,B]\,e^{-2\pi i\int(A,B)}.
\end{split}
\end{align}
This is nothing but the igSPT partition function \eqref{effgauge_Z} with \(B=\hat{A}\). The symmetry extension is now given by
\begin{equation}
    1\rightarrow A\rightarrow \Gamma\times B \rightarrow G\times B\rightarrow 1.
\end{equation}
In \cite{Li:2023knf}, they constructed the lattice model when \(G=B=\mathbb{Z}_2,\Gamma=\mathbb{Z}_4\) and restrict \(\mathbb{Z}_4^\Gamma\times\mathbb{Z}_2^B\rightarrow \mathbb{Z}_4,\mathbb{Z}_2^G\times\mathbb{Z}_2^B\rightarrow \mathbb{Z}_2\) so that keeping the nontriviality of the extension.


\subsection{Other cases}
So far we only considered cases when the original symmetries are zero-form and non-anomalous. We have various generalizations. The first direction is the cases when original symmetries are still zero-form but anomalous. Specifically, we start with the theory with an anomaly
\begin{equation}
    \int_{X_{d+1}}A\cup g^*\tilde{e}+g^*\nu,\quad \delta\nu=e\cup\tilde{e},
\end{equation}
where \([\tilde{e}]\in H^d(BG,\hat{A})\) and \(\nu\in C^{d+1}(BG,\mathbb{R}/\mathbb{Z})\). According to the construction of anomalous symmetry actions on the lattice (see \eqref{ano_sym_action} of Appendix \ref{sec_construct_ano_ham}), the corresponding \(A\) symmetry action is realized by on-site one. Therefore, we can again gauge it and we obtain 
the theory with an anomaly
\begin{equation}
    \int_{X_{d+1}}\hat{A}\cup g^*e+g^*\nu,
\end{equation}
see \cite{Tachikawa:2017gyf} for the details. When we apply our construction \eqref{effgauge_Z} to this case, \(Z_{\mathsf{D}/A}[\hat{A},G,A;e]\) is still anomalous and the anomaly is the same as the original theory. Note that dual symmetry operators are on-site in this case.

The other subtle case is the original anomalies are not linear for \(A\). To generalize our construction to such cases, We may need to gauge non-onsite symmetries. That is not an essential difficulty and we expect we can always gauge such symmetries.\footnote{For \((1+1)\) dimensions, general treatments are discussed in \cite{Seifnashri:2023dpa}.} On the other hand, to handle the remained \(G\)-defects as a global symmetry in the \(A\) gauged theory, we often need to introduce dynamical fields to the defect surfaces and the \(G\) symmetry remained as a \emph{non-invertible} symmetry \cite{Tachikawa:2017gyf,Kaidi:2021xfk}. It would be interesting to study how our formalism \eqref{effgauge_Z} matches with such exotic symmetries.
 
For higher-form or higher-group symmetries, we have similar problems, whereas we may reach similar partition function expressions \eqref{effgauge_Z} for wide classes. Note that no systematic ways to extract algebraic data of anomalies for these symmetries on the lattice have been known yet.

\subsection{Fermionic theory}\label{sec_fermionQFT}
\subsubsection{Fermionization of \(\mathbb{Z}_2\) symmetry}
Let \(\mathsf{D}\) be a \((1+1)d\) bosonic system with a non-anomalous global \(\mathbb{Z}_2\) symmetry. \(Z_\mathsf{D}[A]\) denotes the partition function of \(\mathsf{D}\) with \(\mathbb{Z}_2\) background gauge field. Fermionization of the theory \(\mathsf{D}\) is defined by
\begin{equation}
    Z_{\mathsf{F}}[\rho]=\#\sum_{a}Z_{\mathsf{D}}[a]\,z(\rho,a),
\end{equation}
where \(\rho\in C^1(X_{2},\mathbb{Z}_2)\) indicates the spin structure of spacetime manifolds \(X_2\), i.e. \(\delta\rho=w_2\). Here we introduced the factor \(z(\rho,a)\), which satisfies the following property:
\begin{equation}\label{quad_refin}
    z(\rho,a+a^\prime)
    =z(\rho,a)z(\rho,a^\prime)(-1)^{\int a\cup a^\prime}.
\end{equation}
Namely, \(z(a)\) is a quadratic refinement of the cup product. For \((1+1)d\), we have an explicit expression of \(z(\rho,a)\) \cite{Karch:2019lnn,Ji:2019ugf}:
\begin{equation}
    z(\rho,a)=(-1)^{\operatorname{Arf}(a+\rho)+\operatorname{Arf}(\rho)},
\end{equation}
where \(\operatorname{Arf}\) is the Arf invariant of two-dimensional manifolds.
Indeed, we see that
\begin{align}
\begin{split}
    z(\rho,a+a^\prime)
    &=(-1)^{\operatorname{Arf}((a+a^\prime)+\rho)+\operatorname{Arf}(\rho)}\\
    &=(-1)^{\operatorname{Arf}(a+\rho)+\operatorname{Arf}(a^\prime+\rho)+\operatorname{Arf}(\rho)+\int a\cup a^\prime}(-1)^{\operatorname{Arf}(\rho)}\\
    &=z(\rho,a)z(\rho,a^\prime)(-1)^{\int a\cup a^\prime},
\end{split}
\end{align}
which is a desired property. Here we used the identity for the Arf invariant:
\begin{equation}
    (-1)^{\operatorname{Arf}((a+a^\prime)+\rho)}
    =(-1)^{\operatorname{Arf}(a+\rho)+\operatorname{Arf}(a^\prime+\rho)+\operatorname{Arf}(\rho)+\int a\cup a^\prime}.
\end{equation}
It is also convenient to write \(Z_{\mathsf{F}}[\rho]\) with \(\mathbb{Z}_2\) background field \(S\) for \(\mathbb{Z}_2^F\) (fermion parity symmetry):
\begin{equation}
    Z_{\mathsf{F}}[S+\rho]=\# \sum_{a}Z_{\mathsf{D}}[a]\,(-1)^{\operatorname{Arf}(a+\rho)+\operatorname{Arf}(\rho)+\int a\cup S}.
\end{equation}
We have an inverse operation of fermionization, called bosonization. Bosonization of the theory \(\mathsf{F}\) is given as 
\begin{equation}
    Z_{\mathsf{D}}[A]=\#\sum_{s}Z_{\mathsf{F}}[s+\rho]\,(-1)^{\operatorname{Arf}(A+\rho)+\operatorname{Arf}(\rho)+\int s\cup A}.
\end{equation}
For higher dimensions, \(z(a,\rho)\) was given in \cite{Gu:2012ib,Gaiotto:2015zta}.

\subsubsection{Fermionization of \(\mathbb{Z}_2\) subgroups}\label{sec_fer_subgp}
In the following, we only consider \((1+1)\) dimensions. We want to discuss the fermionization of non-anomalous \(\mathbb{Z}_2\) subgroups. However, when the \(\mathbb{Z}_2\) symmetry is a nontrivial central subgroup of some larger groups, the definition of \(z(\rho,a)\) becomes subtle because \(a\) is not a cocycle in such cases. To overcome this difficulty, we instead define the fermionization as
\begin{equation}\label{bulk_fermionize}
    Z_{\mathsf{F}}[S+\rho,G]=\sum_{\delta a=-g^*e_2}Z_{\mathsf{D}}[G,a]\,z(\rho,a;b)(-1)^{\int a\cup S},
\end{equation}
where \(z(\rho,a;b)\) is given by
\begin{equation}
    z(\rho,a;b)=(-1)^{\int_{X_3} \tilde{\rho}\cup g^*e}\sigma(a;b).
\end{equation}
Here, \(\tilde{\rho}\) is an extension of the spin structure to the three-dimensional manifold \(X_3\) and \(\sigma(a;b)\) is the so-called bulk-boundary Gu-Wen Grassmann integral. This quantity was discussed in \cite{Thorngren:2018bhj} and later constructed more explicitly in \cite{Kobayashi:2019lep}.\footnote{The author thanks Ryohei Kobayashi for suggesting to use the bulk-boundary Gu-Wen Grassmann integral.} \(\sigma(a;b)\) is defined as a \((2+1)d\) bulk theory, and two gauge fields \(a,b\) satisfy \(\delta a=b\). When \(b=0\), \(z(\rho,a;b)\) becomes \(z(\rho,a)\). As shown in \cite{Kobayashi:2019lep}, \(\sigma(a;b)\) transforms with
\begin{equation}
    \sigma(a+\delta\chi;b)=\sigma(a;b)(-1)^{\int\chi\delta a}
\end{equation}
for \(\chi\in C^0(X_3,\mathbb{Z}_2)\).

Using \eqref{bulk_fermionize}, let us implement the fermionization for \(\mathbb{Z}_2\) subgroups. When the theory \(\mathsf{D}\) is non-anomalous, the right-hand side of \eqref{bulk_fermionize} transforms with 
\begin{equation}
    \sum_{a}Z_{\mathsf{D}}[G,a+\delta\chi]\,z(\rho,a+\delta\chi;b)(-1)^{\int (a+\delta\chi)\cup S}=(-1)^{\int \chi(g^*e+\delta S)}\sum_{a}Z_{\mathsf{D}}[G,a]\,z(\rho,a;b)(-1)^{\int a\cup S}
\end{equation}
under the gauge transformation \(a\mapsto a+\delta\chi\). Therefore, the \(\mathbb{Z}_2\) gauge field for \(\mathbb{Z}_2^F\) symmetry should satisfy \(\delta S=g^*e\), which means that the total global symmetry in the fermionized theory is extended by \(\mathbb{Z}_2^F\) symmetry and we have a twisted spin structure. Note that the \(G\) symmetry in the fermionized theory is non-anomalous because \(\mathsf{D}\) is non-anomalous. Precisely, due to the extension by \(\mathbb{Z}_2^F\), we have the bulk Gu-Wen action described by the pair \((\nu,[n])\in C^3(BG,\mathbb{R}/\mathbb{Z})\times H^2(BG,\mathbb{Z}_2)\) with the Gu-Wen equation \(\delta\nu=\frac{1}{2}(n\cup n+e\cup n),\nu=0\). However, since we have the relation \(\delta S=g^*e\), the bulk anomaly inflow is trivialized by boundary counterterms.

When \(\mathsf{D}\) has an anomaly of the form \(2\pi i\int A\cup g^*e+g^*\nu\), the additional phase factor now becomes
\begin{equation}
    (-1)^{\int \chi(g^*e+g^*e+\delta S)},
\end{equation}
and so \(\delta S=0\).
It is now clear that \(3d\) dependence of \(Z_{\mathsf{F}}\) comes from \(z(\rho,a;e^2)\,e^{2\pi i\int g^*\nu}\) and this is nothing but the fermion SPT phase called the Gu-Wen phase \cite{Gu:2012ib}, which is specified by the Gu-Wen data \((\nu,n)\) with the Gu-Wen equation \(\delta\nu=\frac{1}{2}e\cup e\).

\paragraph{Example.} Let us consider the case when \(G=\mathbb{Z}_2\). The full \(\mathbb{Z}_4\) symmetry of the bosonic system is specified by two \(\mathbb{Z}_2\) gauge field \(A,G\) and the extension class \(g^*e\in H^2(X_2,\mathbb{Z}_2)\). Suppose that the system has an anomaly with the following three-cocycle.
\begin{align}\label{Gu-Wen_seed}
    \begin{split}
        \omega&=\int A\cup g^*e+g^*\nu\\
        &=\frac{1}{2}\int A\cup G^2+\frac{1}{2}G^3.
    \end{split}
\end{align}
This corresponds to \(2\mod 4\) in \(H^3(B\mathbb{Z}_4,U(1))\cong\mathbb{Z}_4\).
When we fermionize the \(\mathbb{Z}_2^A\) symmetry, we obtain the nontrivial Gu-Wen \(\mathbb{Z}_2\) anomaly.

In Sec.~\ref{sec_fermion_gapless}, we have constructed the lattice model for the fermionic igSPT phase. In the construction, we started with a non-anomalous \(\mathbb{Z}_8\) symmetry and gauge the non-anomalous \(\mathbb{Z}_4\) symmetry. Let us see the meaning of the construction. When we gauge the \(\mathbb{Z}_4\) non-anomalous subgroup of a \(\mathbb{Z}_8\) symmetry, the system has a mixed anomaly between \(\mathbb{Z}_4\) and \(\mathbb{Z}_2\), which is specified by 
\begin{equation}
    \frac{1}{4}\int A^\prime\cup (g^\prime)^*e,
\end{equation}
where \(A^\prime\) is a dual \(\mathbb{Z}_4\) gauge field and \(e\) specifies the extension class of \(1\rightarrow\mathbb{Z}_4\rightarrow\mathbb{Z}_8\rightarrow\mathbb{Z}_2\rightarrow 1\). Then we consider a \(\mathbb{Z}_4\) subgroup of \(\mathbb{Z}_4\times\mathbb{Z}_2\), which is generated by two \(\mathbb{Z}_2\) gauge fields \(A\) and \(G\) and they are chosen so that 
\begin{equation}
    A^\prime=2A+G,\quad G^\prime=G
\end{equation}
We can see this \(\mathbb{Z}_4\) subgroup corresponds to the \(\mathbb{Z}_4\) symmetry \eqref{Z4_condense} which we used in Sec.~\ref{sec_fermion_gapless}. Since \((g^\prime)^*e=G^\prime\cup G^\prime\), the anomaly three-cocycle of this \(\mathbb{Z}_4\) subgroup is 
\begin{equation}
    \frac{1}{4}\int\left(2A+G\right)G^2=\frac{1}{2}\int A\cup G^2+\frac{1}{2}G^3,
\end{equation}
which is equal to \eqref{Gu-Wen_seed}.

\subsubsection{Our construction}
We are interested in the lattice model \eqref{gaugeHamF} or \eqref{fer_effgauge}. Our claim is almost parallel to the bosonic case. The corresponding partition function is given by 
\begin{equation}\label{eff_gaugeFZ}
    Z_{\mathsf{F}}[\rho,G,A;e]=\#\sum_{a}Z_{\mathsf{D}}[G,a;e]\,z(\rho,a+A;e).
\end{equation}
The difference to the bosonic case \eqref{effgauge_Z} is the \(\mathbb{Z}_2\) background field \(A\) is not decoupled to the \(\mathbb{Z}_2\) dynamical field \(a\). Therefore there is no unified interpretation of unitary conjugation by \(V_f\) \eqref{Vf}.

\paragraph{Example.} Let us see some examples when \(G\) is trivial. For bosonic theory \(\mathsf{D}\), we have two gapped phases \(Z_{\mathsf{D}}[A]=1,\#\delta(A)\). When \(Z_{\mathsf{D}}[A]=1\), \eqref{eff_gaugeFZ} is 
\begin{equation}
    Z_{\mathsf{F}}[\rho,0,A;0]=\#\sum_{a}(-1)^{\operatorname{Arf}(a+A+\rho)+\operatorname{Arf}(\rho)}=\#(-1)^{\operatorname{Arf}(\rho)},
\end{equation}
which describes the partition function of the Kitaev phase. When \(Z_{\mathsf{D}}[A]=\#\delta(A)\), we obtain 
\begin{equation}
    Z_{\mathsf{F}}[\rho,0,A;0]=\#\sum_{a}\delta(a)(-1)^{\operatorname{Arf}(a+A+\rho)+\operatorname{Arf}(\rho)}=\#(-1)^{\operatorname{Arf}(A+\rho)+\operatorname{Arf}(\rho)},
\end{equation}
which describes the nontrivial fermionic \(\mathbb{Z}_2\) SPT phase. The calculation here is consistent with the lattice model discussions in Sec.~\ref{sec_non-ano_fermion}.

\subsubsection{Generalizations}
For \(d\) dimensions, bosonizing the Gu-Wen fermion anomalous system yields a \((d-2)\)-group symmetry \cite{Thorngren:2018bhj}, and the anomaly of the bosonized system is a mixture between a \((d-2)\)-form \(\mathbb{Z}_2\) symmetry and a zero-form \(G\) symmetry. For more general fermion anomalies, one can use higher bosonizations \cite{Thorngren:2018bhj}. Algebraic data of symmetries in higher bosonized theories are complicated in general. To illustrate, the bosonic shadow of the generator of \((2+1)d\) unitary \(\mathbb{Z}_2\) fermionic SPT phases are realized by the toric code phase with the \(em\) exchange \(\mathbb{Z}_2\) symmetry \cite{Bhardwaj:2016clt}, and the \(\mathbb{Z}_2\) symmetry acts on \((1+1)d\) boundary systems as Kramers-Wannier duality symmetry, which is non-invertible \cite{Jones:2019lwm}. We expect we can generalize formalisms in this note to such exotic anomalies and leave them for future work.

\section*{Acknowledgements}
The author thanks Kansei Inamura, Ryohei Kobayashi, Shuhei Ohyama, and Shinsei Ryu for useful discussions on this work. The author thanks Yunqin Zheng for insightful discussions on the early stage of this work. The author thanks Ryohei Kobayashi, Shuhei Ohyama, Masatoshi Sato, Ken Shiozaki, and Yuya Tanizaki for comments on a draft.

\emph{Note added.} A part of the present work was reported in \cite{Ando:2023jps}. During the completion of this note, the author became aware of a related paper by Lei Su and Meng Zeng \cite{Su:2024vrk}, which constructs gapless SPT models by gauging non-anomalous subgroups. The contents have some overlap with our result, in particular with Sec.~\ref{sec_gapless_boson}.

\begin{appendices}
\section{Construction of anomalous lattice models}\label{sec_construct_ano_ham}
Consider a spatial \(d\)-dimensional lattice system. Suppose that this system has an 't Hooft anomaly with zero-form \(G\) symmetry, and take a (nontrivial) cocycle element \(\omega\in H^{d+2}(G,U(1))\) for this anomaly.

Let \(\mathcal{H}_{j}\) be a local Hilbert space defined on a vertex \(j\). The total Hilbert space \(\mathcal{H}\) is \(\bigotimes_{j}\mathcal{H}_{j}\). We write the base of each vertex Hilbert space \(\ket{g}_{j}\), and we fix them. Anomalous symmetry action on \(\ket{\{g_{j}\}}\) are given as follows \cite{Else:2014vma,Wang:2017loc}.
\begin{equation}\label{ano_sym_action}
    U_\omega(g)\ket{\{g_{j}\}}=\prod_{(i_{0},\ldots,i_{d})}
    \omega(g_{i_0}^{-1}g_{i_1},g_{i_1}^{-1}g_{i_2},\ldots,g_{i_d}^{-1}g^{-1},g)^{s(i_{0},\ldots,i_{d})}
    \ket{\{gg_{j}\}},
\end{equation}
where \((i_{0},\ldots,i_{d})\) is a \(d\)-simplex that lives in \(d\)-dimensional space, and \(s(i_{0},\ldots,i_{d})\) is a sign of \((i_{0},\ldots,i_{d})\). To construct an anomalous Hamiltonian, we first prepare a trivial Hamiltonian:
\begin{equation}\label{ano_seed_ham}
    H_0=-\sum_{j}h^0_j, \quad
    h^0_j=\frac{1}{\sqrt{{|G|}}}\sum_{g\in G}\ket{g}_j\bra{g}_j
\end{equation}
From this, we can construct anomalous Hamiltonians by symmetrizing \eqref{ano_seed_ham} as follows:
\begin{equation}
    H_\omega=\sum_{g\in G}U_\omega(g)H_0 U_\omega(g)^\dagger.
\end{equation}
Note that the concrete expression of \(H_\omega\) depends on the choice of the Hamiltonian \eqref{ano_seed_ham}.

\end{appendices}

\bibliographystyle{ytamsalpha}
\bibliography{ref}

\newcommand{\etalchar}[1]{$^{#1}$}
\providecommand{\bysame}{\leavevmode\hbox to3em{\hrulefill}\thinspace}
\providecommand{\MR}{\relax\ifhmode\unskip\space\fi MR }
\providecommand{\MRhref}[2]{%
  \href{http://www.ams.org/mathscinet-getitem?mr=#1}{#2}
}
\providecommand{\href}[2]{#2}
\providecommand{\doihref}[2]{\href{#1}{#2}}
\providecommand{\arxivfont}{\tt}
\begin{thebibliography}{CGLW11}

\bibitem[And23]{Ando:2023jps}
T.~Ando, \emph{Duality constructions of topological quantum critical models},
  Japan Physics Society 2023 Annual (78th) Meeting, 18aB203-9: unpublished, 9
  2023.

\bibitem[BGK16]{Bhardwaj:2016clt}
L.~Bhardwaj, D.~Gaiotto, and A.~Kapustin, \emph{{State sum constructions of
  spin-TFTs and string net constructions of fermionic phases of matter}},
  \doihref{http://dx.doi.org/10.1007/JHEP04(2017)096}{JHEP \textbf{04} (2017)
  096}, \href{http://arxiv.org/abs/1605.01640}{{\arxivfont arXiv:1605.01640
  [cond-mat.str-el]}}.

\bibitem[BVSM20]{Borla:2020avq}
U.~Borla, R.~Verresen, J.~Shah, and S.~Moroz, \emph{{Gauging the Kitaev
  chain}}, \doihref{http://dx.doi.org/10.21468/SciPostPhys.10.6.148}{SciPost
  Phys. \textbf{10} (2021) 148},
  \href{http://arxiv.org/abs/2010.00607}{{\arxivfont arXiv:2010.00607
  [cond-mat.str-el]}}.

\bibitem[CGLW11]{Chen:2011pg}
X.~Chen, Z.-C. Gu, Z.-X. Liu, and X.-G. Wen, \emph{{Symmetry protected
  topological orders and the group cohomology of their symmetry group}},
  \doihref{http://dx.doi.org/10.1103/PhysRevB.87.155114}{Phys. Rev. B
  \textbf{87} (2013) 155114}, \href{http://arxiv.org/abs/1106.4772}{{\arxivfont
  arXiv:1106.4772 [cond-mat.str-el]}}.

\bibitem[CGW10]{Chen:2010zpc}
X.~Chen, Z.-C. Gu, and X.-G. Wen, \emph{{Classification of gapped symmetric
  phases in one-dimensional spin systems}},
  \doihref{http://dx.doi.org/10.1103/PhysRevB.83.035107}{Phys. Rev. B
  \textbf{83} (2011) 035107}, \href{http://arxiv.org/abs/1008.3745}{{\arxivfont
  arXiv:1008.3745 [cond-mat.str-el]}}.

\bibitem[Che19]{Chen:2019wlx}
Y.-A. Chen, \emph{{Exact bosonization in arbitrary dimensions}},
  \doihref{http://dx.doi.org/10.1103/PhysRevResearch.2.033527}{Phys. Rev. Res.
  \textbf{2} (2020) 033527}, \href{http://arxiv.org/abs/1911.00017}{{\arxivfont
  arXiv:1911.00017 [cond-mat.str-el]}}.

\bibitem[CK18]{Chen:2018nog}
Y.-A. Chen and A.~Kapustin, \emph{{Bosonization in three spatial dimensions and
  a 2-form gauge theory}},
  \doihref{http://dx.doi.org/10.1103/PhysRevB.100.245127}{Phys. Rev. B
  \textbf{100} (2019) 245127},
  \href{http://arxiv.org/abs/1807.07081}{{\arxivfont arXiv:1807.07081
  [cond-mat.str-el]}}.

\bibitem[CKR17]{Chen:2017fvr}
Y.-A. Chen, A.~Kapustin, and D.~Radi\v{c}evi\'c, \emph{{Exact bosonization in
  two spatial dimensions and a new class of lattice gauge theories}},
  \doihref{http://dx.doi.org/10.1016/j.aop.2018.03.024}{Annals Phys.
  \textbf{393} (2018) 234--253},
  \href{http://arxiv.org/abs/1711.00515}{{\arxivfont arXiv:1711.00515
  [cond-mat.str-el]}}.

\bibitem[CLV13]{Chen:2014zvm}
X.~Chen, Y.-M. Lu, and A.~Vishwanath, \emph{{Symmetry-protected topological
  phases from decorated domain walls}},
  \doihref{http://dx.doi.org/10.1038/ncomms4507}{Nature Commun. \textbf{5}
  (2014) 3507}, \href{http://arxiv.org/abs/1303.4301}{{\arxivfont
  arXiv:1303.4301 [cond-mat.str-el]}}.

\bibitem[EN14]{Else:2014vma}
D.~V. Else and C.~Nayak, \emph{{Classifying symmetry-protected topological
  phases through the anomalous action of the symmetry on the edge}},
  \doihref{http://dx.doi.org/10.1103/PhysRevB.90.235137}{Phys. Rev. B
  \textbf{90} (2014) 235137}, \href{http://arxiv.org/abs/1409.5436}{{\arxivfont
  arXiv:1409.5436 [cond-mat.str-el]}}.

\bibitem[GK15]{Gaiotto:2015zta}
D.~Gaiotto and A.~Kapustin, \emph{{Spin TQFTs and fermionic phases of matter}},
  \doihref{http://dx.doi.org/10.1142/S0217751X16450445}{Int. J. Mod. Phys. A
  \textbf{31} (2016) 1645044},
  \href{http://arxiv.org/abs/1505.05856}{{\arxivfont arXiv:1505.05856
  [cond-mat.str-el]}}.

\bibitem[GW12]{Gu:2012ib}
Z.-C. Gu and X.-G. Wen, \emph{{Symmetry-protected topological orders for
  interacting fermions: Fermionic topological nonlinear \ensuremath{\sigma}
  models and a special group supercohomology theory}},
  \doihref{http://dx.doi.org/10.1103/PhysRevB.90.115141}{Phys. Rev. B
  \textbf{90} (2014) 115141}, \href{http://arxiv.org/abs/1201.2648}{{\arxivfont
  arXiv:1201.2648 [cond-mat.str-el]}}.

\bibitem[HC23]{Huang:2023pyk}
S.-J. Huang and M.~Cheng, \emph{{Topological holography, quantum criticality,
  and boundary states}},
  \doihref{http://dx.doi.org/10.21468/SciPostPhys.18.6.213}{SciPost Phys.
  \textbf{18} (2025) 213}, \href{http://arxiv.org/abs/2310.16878}{{\arxivfont
  arXiv:2310.16878 [cond-mat.str-el]}}.

\bibitem[HFUT22]{Hidaka:2022bbz}
Y.~Hidaka, S.~C. Furuya, A.~Ueda, and Y.~Tada, \emph{{Gapless
  symmetry-protected topological phase of quantum antiferromagnets on
  anisotropic triangular strip}},
  \doihref{http://dx.doi.org/10.1103/PhysRevB.106.144436}{Phys. Rev. B
  \textbf{106} (2022) 144436},
  \href{http://arxiv.org/abs/2205.15525}{{\arxivfont arXiv:2205.15525
  [cond-mat.str-el]}}.

\bibitem[HNT20]{Hsieh:2020uwb}
C.-T. Hsieh, Y.~Nakayama, and Y.~Tachikawa, \emph{{Fermionic minimal models}},
  \doihref{http://dx.doi.org/10.1103/PhysRevLett.126.195701}{Phys. Rev. Lett.
  \textbf{126} (2021) 195701},
  \href{http://arxiv.org/abs/2002.12283}{{\arxivfont arXiv:2002.12283
  [cond-mat.str-el]}}.

\bibitem[Ina22]{Inamura:2022lun}
K.~Inamura, \emph{{Fermionization of fusion category symmetries in 1+1
  dimensions}}, \doihref{http://dx.doi.org/10.1007/JHEP10(2023)101}{JHEP
  \textbf{10} (2023) 101}, \href{http://arxiv.org/abs/2206.13159}{{\arxivfont
  arXiv:2206.13159 [cond-mat.str-el]}}.

\bibitem[JM19]{Jones:2019lwm}
R.~A. Jones and M.~A. Metlitski, \emph{{One-dimensional lattice models for the
  boundary of two-dimensional Majorana fermion symmetry-protected topological
  phases: Kramers-Wannier duality as an exact Z2 symmetry}},
  \doihref{http://dx.doi.org/10.1103/PhysRevB.104.245130}{Phys. Rev. B
  \textbf{104} (2021) 245130},
  \href{http://arxiv.org/abs/1902.05957}{{\arxivfont arXiv:1902.05957
  [cond-mat.str-el]}}.

\bibitem[JSW19]{Ji:2019ugf}
W.~Ji, S.-H. Shao, and X.-G. Wen, \emph{{Topological Transition on the
  Conformal Manifold}},
  \doihref{http://dx.doi.org/10.1103/PhysRevResearch.2.033317}{Phys. Rev. Res.
  \textbf{2} (2020) 033317}, \href{http://arxiv.org/abs/1909.01425}{{\arxivfont
  arXiv:1909.01425 [cond-mat.str-el]}}.

\bibitem[Kit00]{Kitaev:2000nmw}
A.~Kitaev, \emph{{Unpaired Majorana fermions in quantum wires}},
  \doihref{http://dx.doi.org/10.1070/1063-7869/44/10S/S29}{Phys. Usp.
  \textbf{44} (2001) 131--136},
  \href{http://arxiv.org/abs/cond-mat/0010440}{{\arxivfont
  arXiv:cond-mat/0010440}}.

\bibitem[KL21]{Kawagoe:2021gqi}
K.~Kawagoe and M.~Levin, \emph{{Anomalies in bosonic symmetry-protected
  topological edge theories: Connection to F symbols and a method of
  calculation}}, \doihref{http://dx.doi.org/10.1103/PhysRevB.104.115156}{Phys.
  Rev. B \textbf{104} (2021) 115156},
  \href{http://arxiv.org/abs/2105.02909}{{\arxivfont arXiv:2105.02909
  [cond-mat.str-el]}}.

\bibitem[KOT19]{Kobayashi:2019lep}
R.~Kobayashi, K.~Ohmori, and Y.~Tachikawa, \emph{{On gapped boundaries for SPT
  phases beyond group cohomology}},
  \doihref{http://dx.doi.org/10.1007/JHEP11(2019)131}{JHEP \textbf{11} (2019)
  131}, \href{http://arxiv.org/abs/1905.05391}{{\arxivfont arXiv:1905.05391
  [cond-mat.str-el]}}.

\bibitem[KOZ21]{Kaidi:2021xfk}
J.~Kaidi, K.~Ohmori, and Y.~Zheng, \emph{{Kramers-Wannier-like Duality Defects
  in (3+1)D Gauge Theories}},
  \doihref{http://dx.doi.org/10.1103/PhysRevLett.128.111601}{Phys. Rev. Lett.
  \textbf{128} (2022) 111601},
  \href{http://arxiv.org/abs/2111.01141}{{\arxivfont arXiv:2111.01141
  [hep-th]}}.

\bibitem[KT17]{Kapustin:2017jrc}
A.~Kapustin and R.~Thorngren, \emph{{Fermionic SPT phases in higher dimensions
  and bosonization}}, \doihref{http://dx.doi.org/10.1007/JHEP10(2017)080}{JHEP
  \textbf{10} (2017) 080}, \href{http://arxiv.org/abs/1701.08264}{{\arxivfont
  arXiv:1701.08264 [cond-mat.str-el]}}.

\bibitem[KTT19]{Karch:2019lnn}
A.~Karch, D.~Tong, and C.~Turner, \emph{{A Web of 2d Dualities: ${\bf Z}_2$
  Gauge Fields and Arf Invariants}},
  \doihref{http://dx.doi.org/10.21468/SciPostPhys.7.1.007}{SciPost Phys.
  \textbf{7} (2019) 007}, \href{http://arxiv.org/abs/1902.05550}{{\arxivfont
  arXiv:1902.05550 [hep-th]}}.

\bibitem[LG12]{Levin:2012yb}
M.~Levin and Z.-C. Gu, \emph{{Braiding statistics approach to
  symmetry-protected topological phases}},
  \doihref{http://dx.doi.org/10.1103/PhysRevB.86.115109}{Phys. Rev. B
  \textbf{86} (2012) 115109}, \href{http://arxiv.org/abs/1202.3120}{{\arxivfont
  arXiv:1202.3120 [cond-mat.str-el]}}.

\bibitem[LOZ22]{Li:2022jbf}
L.~Li, M.~Oshikawa, and Y.~Zheng, \emph{{Decorated defect construction of
  gapless-SPT states}},
  \doihref{http://dx.doi.org/10.21468/SciPostPhys.17.1.013}{SciPost Phys.
  \textbf{17} (2024) 013}, \href{http://arxiv.org/abs/2204.03131}{{\arxivfont
  arXiv:2204.03131 [cond-mat.str-el]}}.

\bibitem[LOZ23]{Li:2023knf}
\bysame, \emph{{Intrinsically/purely gapless-SPT from non-invertible duality
  transformations}},
  \doihref{http://dx.doi.org/10.21468/SciPostPhys.18.5.153}{SciPost Phys.
  \textbf{18} (2025) 153}, \href{http://arxiv.org/abs/2307.04788}{{\arxivfont
  arXiv:2307.04788 [cond-mat.str-el]}}.

\bibitem[LYTC14]{Li:2014pta}
W.~Li, S.~Yang, H.-H. Tu, and M.~Cheng, \emph{{Criticality in
  Translation-Invariant Parafermion Chains}},
  \doihref{http://dx.doi.org/10.1103/PhysRevB.91.115133}{Phys. Rev. B
  \textbf{91} (2015) 115133}, \href{http://arxiv.org/abs/1407.3790}{{\arxivfont
  arXiv:1407.3790 [cond-mat.str-el]}}.

\bibitem[MZW21]{Ma:2021dfx}
R.~Ma, L.~Zou, and C.~Wang, \emph{{Edge physics at the deconfined transition
  between a quantum spin Hall insulator and a superconductor}},
  \doihref{http://dx.doi.org/10.21468/SciPostPhys.12.6.196}{SciPost Phys.
  \textbf{12} (2022) 196}, \href{http://arxiv.org/abs/2110.08280}{{\arxivfont
  arXiv:2110.08280 [cond-mat.str-el]}}.

\bibitem[PBTO09]{Pollmann:2009mhk}
F.~Pollmann, E.~Berg, A.~M. Turner, and M.~Oshikawa, \emph{{Symmetry protection
  of topological phases in one-dimensional quantum spin systems}},
  \doihref{http://dx.doi.org/10.1103/PhysRevB.85.075125}{Phys. Rev. B
  \textbf{85} (2012) 075125}, \href{http://arxiv.org/abs/0909.4059}{{\arxivfont
  arXiv:0909.4059 [cond-mat.str-el]}}.

\bibitem[PTBO09]{Pollmann:2009ryx}
F.~Pollmann, A.~M. Turner, E.~Berg, and M.~Oshikawa, \emph{{Entanglement
  spectrum of a topological phase in one dimension}},
  \doihref{http://dx.doi.org/10.1103/PhysRevB.81.064439}{Phys. Rev. B
  \textbf{81} (2010) 064439}, \href{http://arxiv.org/abs/0910.1811}{{\arxivfont
  arXiv:0910.1811 [cond-mat.str-el]}}.

\bibitem[RB01]{Raussendorf:2001zim}
R.~Raussendorf and H.~J. Briegel, \emph{{A One-Way Quantum Computer}},
  \doihref{http://dx.doi.org/10.1103/PhysRevLett.86.5188}{Phys. Rev. Lett.
  \textbf{86} (2001) 5188}.

\bibitem[Sei23]{Seifnashri:2023dpa}
S.~Seifnashri, \emph{{Lieb-Schultz-Mattis anomalies as obstructions to gauging
  (non-on-site) symmetries}},
  \doihref{http://dx.doi.org/10.21468/SciPostPhys.16.4.098}{SciPost Phys.
  \textbf{16} (2024) 098}, \href{http://arxiv.org/abs/2308.05151}{{\arxivfont
  arXiv:2308.05151 [cond-mat.str-el]}}.

\bibitem[SPGC10]{Schuch:2010}
N.~Schuch, D.~Pérez-García, and I.~Cirac, \emph{{Classifying quantum phases
  using matrix product states and projected entangled pair states}},
  \doihref{http://dx.doi.org/10.1103/PhysRevB.84.165139}{Phys. Rev. B
  \textbf{84} (2011) 165139}, \href{http://arxiv.org/abs/1010.3732}{{\arxivfont
  arXiv:1010.3732 [cond-mat.str-el]}}.

\bibitem[SPV17]{Scaffidi:2017ppg}
T.~Scaffidi, D.~E. Parker, and R.~Vasseur, \emph{{Gapless Symmetry Protected
  Topological Order}},
  \doihref{http://dx.doi.org/10.1103/PhysRevX.7.041048}{Phys. Rev. X \textbf{7}
  (2017) 041048}, \href{http://arxiv.org/abs/1705.01557}{{\arxivfont
  arXiv:1705.01557 [cond-mat.str-el]}}.

\bibitem[Su23]{Su:2023hud}
L.~Su, \emph{{Boundary criticality via gauging finite subgroups: a case study
  on the clock model}}, \href{http://arxiv.org/abs/2306.02976}{{\arxivfont
  arXiv:2306.02976 [cond-mat.str-el]}}.

\bibitem[SZ24]{Su:2024vrk}
L.~Su and M.~Zeng, \emph{{Gapless symmetry protected topological phases and
  generalized deconfined critical points from gauging a finite subgroup}},
  \href{http://arxiv.org/abs/2401.11702}{{\arxivfont arXiv:2401.11702
  [cond-mat.str-el]}}.

\bibitem[Tac17]{Tachikawa:2017gyf}
Y.~Tachikawa, \emph{{On gauging finite subgroups}},
  \doihref{http://dx.doi.org/10.21468/SciPostPhys.8.1.015}{SciPost Phys.
  \textbf{8} (2020) 015}, \href{http://arxiv.org/abs/1712.09542}{{\arxivfont
  arXiv:1712.09542 [hep-th]}}.

\bibitem[Tho18]{Thorngren:2018bhj}
R.~Thorngren, \emph{{Anomalies and Bosonization}},
  \doihref{http://dx.doi.org/10.1007/s00220-020-03830-0}{Commun. Math. Phys.
  \textbf{378} (2020) 1775--1816},
  \href{http://arxiv.org/abs/1810.04414}{{\arxivfont arXiv:1810.04414
  [cond-mat.str-el]}}.

\bibitem[TVV20]{Thorngren:2020wet}
R.~Thorngren, A.~Vishwanath, and R.~Verresen, \emph{{Intrinsically gapless
  topological phases}},
  \doihref{http://dx.doi.org/10.1103/PhysRevB.104.075132}{Phys. Rev. B
  \textbf{104} (2021) 075132},
  \href{http://arxiv.org/abs/2008.06638}{{\arxivfont arXiv:2008.06638
  [cond-mat.str-el]}}.

\bibitem[VBV{\etalchar{+}}22]{Verresen:2022mcr}
R.~Verresen, U.~Borla, A.~Vishwanath, S.~Moroz, and R.~Thorngren, \emph{{Higgs
  Condensates are Symmetry-Protected Topological Phases: I. Discrete
  Symmetries}}, \href{http://arxiv.org/abs/2211.01376}{{\arxivfont
  arXiv:2211.01376 [cond-mat.str-el]}}.

\bibitem[VTJP19]{Verresen:2019igf}
R.~Verresen, R.~Thorngren, N.~G. Jones, and F.~Pollmann, \emph{{Gapless
  Topological Phases and Symmetry-Enriched Quantum Criticality}},
  \doihref{http://dx.doi.org/10.1103/PhysRevX.11.041059}{Phys. Rev. X
  \textbf{11} (2021) 041059},
  \href{http://arxiv.org/abs/1905.06969}{{\arxivfont arXiv:1905.06969
  [cond-mat.str-el]}}.

\bibitem[WP22]{Wen:2022tkg}
R.~Wen and A.~C. Potter, \emph{{Bulk-boundary correspondence for intrinsically
  gapless symmetry-protected topological phases from group cohomology}},
  \doihref{http://dx.doi.org/10.1103/PhysRevB.107.245127}{Phys. Rev. B
  \textbf{107} (2023) 245127},
  \href{http://arxiv.org/abs/2208.09001}{{\arxivfont arXiv:2208.09001
  [cond-mat.str-el]}}.

\bibitem[WP23]{Wen:2023otf}
\bysame, \emph{{Classification of 1+1D gapless symmetry protected phases via
  topological holography}},
  \doihref{http://dx.doi.org/10.1103/PhysRevB.111.115161}{Phys. Rev. B
  \textbf{111} (2025) 115161},
  \href{http://arxiv.org/abs/2311.00050}{{\arxivfont arXiv:2311.00050
  [cond-mat.str-el]}}.

\bibitem[WWW17]{Wang:2017loc}
J.~Wang, X.-G. Wen, and E.~Witten, \emph{{Symmetric Gapped Interfaces of SPT
  and SET States: Systematic Constructions}},
  \doihref{http://dx.doi.org/10.1103/PhysRevX.8.031048}{Phys. Rev. X \textbf{8}
  (2018) 031048}, \href{http://arxiv.org/abs/1705.06728}{{\arxivfont
  arXiv:1705.06728 [cond-mat.str-el]}}.

\bibitem[YHS{\etalchar{+}}21]{Yu:2021rng}
X.-J. Yu, R.-Z. Huang, H.-H. Song, L.~Xu, C.~Ding, and L.~Zhang,
  \emph{{Conformal Boundary Conditions of Symmetry-Enriched Quantum Critical
  Spin Chains}},
  \doihref{http://dx.doi.org/10.1103/PhysRevLett.129.210601}{Phys. Rev. Lett.
  \textbf{129} (2022) 210601},
  \href{http://arxiv.org/abs/2111.10945}{{\arxivfont arXiv:2111.10945
  [cond-mat.str-el]}}.

\bibitem[YYLJ24]{Yu:2024toh}
X.-J. Yu, S.~Yang, H.-Q. Lin, and S.-K. Jian, \emph{{Universal Entanglement
  Spectrum in One-Dimensional Gapless Symmetry Protected Topological States}},
  \doihref{http://dx.doi.org/10.1103/PhysRevLett.133.026601}{Phys. Rev. Lett.
  \textbf{133} (2024) 026601},
  \href{http://arxiv.org/abs/2402.04042}{{\arxivfont arXiv:2402.04042
  [cond-mat.str-el]}}.

\bibitem[ZL22]{Zhang:2022wwn}
C.~Zhang and M.~Levin, \emph{{Exactly Solvable Model for a Deconfined Quantum
  Critical Point in 1D}},
  \doihref{http://dx.doi.org/10.1103/PhysRevLett.130.026801}{Phys. Rev. Lett.
  \textbf{130} (2023) 026801},
  \href{http://arxiv.org/abs/2206.01222}{{\arxivfont arXiv:2206.01222
  [cond-mat.str-el]}}.

\end{thebibliography}
\end{document}